\documentclass[a4paper,11pt]{article}
\usepackage{graphicx,amssymb,amstext,amsmath,mathabx,mathtools,esvect}
\usepackage{color}
\usepackage{floatflt}

\definecolor{cbl}{rgb}{0,0,1}                % bleu\

\newcommand{\bc}{\begin{center}}
\newcommand{\ec}{\end{center}}
\def\ba#1{\begin{array}{#1}\displaystyle}
\newcommand{\ea}{\end{array}}

\newcommand{\beq}{\begin{equation}}
\newcommand{\eeq}{\end{equation}}
\newcommand{\beqa}{\begin{eqnarray}}
\newcommand{\eeqa}{\end{eqnarray}}

\newcommand{\bi}{\begin{itemize}}
\newcommand{\ei}{\end{itemize}}
\newcommand{\bra}{\langle}
\newcommand{\ket}{\rangle}

\newcommand{\bal}{\boldsymbol{\alpha}}
\newcommand{\bol}{\boldsymbol{0}}
\newcommand{\bel}{\boldsymbol{\beta}}

\newcommand{\TTb}{\mathrm{T}\overline{\mathrm{T}}}

\usepackage{cite}

\definecolor{purple_nice}{rgb}{0.4,0.2,0.7}
\definecolor{fuel_blue}{RGB}{42,162,185}
\definecolor{YInMn_blue}{RGB}{46, 80, 144}
\definecolor{ultramarine}{RGB}{63, 0, 255}
\definecolor{KLEIN_blue}{rgb}{0, 0.18, 0.65}
\usepackage[linktocpage=true]{hyperref}
\hypersetup{
    colorlinks=true,
    linkcolor=YInMn_blue,
    citecolor=ultramarine,
    filecolor=fuel_blue,
    urlcolor=KLEIN_blue,
}
\makeatletter
\renewenvironment{abstract}{%
      \begin{center}%
        {\bfseries \normalsize\abstractname\vspace{\z@}}%  %% <- here I've added \Large
      \end{center}%
      \quotation}
    {\endquotation}
\makeatother
\usepackage[normalem]{ulem}  % FOR THE \sout COMMAND

\begin{document}

\begin{titlepage}
\title{{ Form Factors and Correlation Functions of $\mathrm{T}\overline{\mathrm{T}}$-Deformed Integrable Quantum Field Theories}}
\author{Olalla A. Castro-Alvaredo${}^{\heartsuit}$, Stefano Negro${}^\clubsuit$ and Fabio Sailis${}^{\diamondsuit}$\\[0.3cm]}
\date{{\small ${}^{\heartsuit,\diamondsuit}$ Department of Mathematics, City, University of London, 10 Northampton Square EC1V 0HB, UK\\[0.3cm]
${}^{\clubsuit}$ Center for Cosmology and Particle Physics, New York University, New York, NY 10003, U.S.A.\\[0.3cm]}}
\maketitle
\begin{abstract}
\normalsize

The study of $\mathrm{T}\overline{\mathrm{T}}$-perturbed quantum field theories is an active area of research with deep connections to fundamental aspects of the scattering theory of integrable quantum field theories, generalised Gibbs ensembles, and string theory. Many features of these theories, such as the peculiar behaviour of their ground state energy and the form of their scattering matrices, have been studied in the literature. However, so far, very few studies have approached these theories from the viewpoint of the form factor program. From the perspective of scattering theory, the effects of a $\mathrm{T}\overline{\mathrm{T}}$ perturbation (and higher spin versions thereof) is encoded in a universal deformation of the two-body scattering matrix by a CDD factor. It is then natural to ask how these perturbations influence the form factor equations and, more generally, the form factor program. In this paper, we address this question for free theories, although some of our results extend more generally. We show that the form factor equations admit general solutions and how these can help us study the distinct behaviour of correlation functions at short distances in theories perturbed by irrelevant operators.\\

\end{abstract}
\vfill

\noindent 
${}^{\heartsuit}$ o.castro-alvaredo@city.ac.uk\\
${}^{\clubsuit}$ stefano.negro@nyu.edu\\
${}^{\diamondsuit}$ fabio.sailis@city.ac.uk\\

\hfill \today
\end{titlepage}
\section{Introduction}

The seminal work of A.B.~Zamoldochikov \cite{Zamolodchikov:1989hfa} proposed a viewpoint of integrable quantum field theories (IQFTs) as massive perturbations of critical points, described by conformal field theory (CFT). These perturbations are relevant and associated with specific fields of the underlying CFT having conformal dimension $\Delta<1$. However, it is also possible to consider irrelevant deformations – of both conformal and gapped quantum field theories. Recently, a particular class of such perturbations has attracted considerable interest: those generated by
the field $\mathrm{T}\overline{\mathrm{T}}$, where $T$ and $\bar{T}$ are the holomorphic and antiholomorphic components of the stress energy tensor. The earliest study of this field’s properties, particularly its vacuum expectation value, was carried out in \cite{Zamolodchikov:2004ce} for generic 2D quantum field theory and quickly followed by a systematic study of the form factors of the $\mathrm{T}\overline{\mathrm{T}}$ operator itself \cite{Delfino:2004vc, Delfino:2006te} in massive IQFTs. 

The works \cite{Smirnov:2016lqw, Cavaglia:2016oda} showed that the properties of a 2D Quantum Field Theory perturbed by the irrelevant composite field $\mathrm{T}\overline{\mathrm{T}}$, which has left and right conformal dimension 2, are under exceptionally good control, even deep in the UV. In particular, one can regard the $\mathrm{T}\overline{\mathrm{T}}$ deformation as being \emph{solvable}, in the sense that physical observables of interest, such as the finite-volume spectrum, the $S$-matrix \cite{Smirnov:2016lqw, Cavaglia:2016oda, Dubovsky:2017cnj} and the partition functions \cite{Dubovsky:2018bmo, Cardy:2018sdv, Datta:2018thy}, can all be determined exactly in terms of the corresponding undeformed quantities. The property of being solvable is also present in the generalised $\mathrm{T}\overline{\mathrm{T}}$ deformations, obtained by perturbing an IQFT by composite operators constructed from higher-spin conserved currents \cite{Conti:2019dxg, Hernandez-Chifflet:2019sua}. Performing a (generalised) $\mathrm{T}\overline{\mathrm{T}}$-deformation in an IQFT is equivalent to modifying the two-body scattering matrix by a particular type of CDD factor \cite{Smirnov:2016lqw, Cavaglia:2016oda, Camilo:2021gro, Cordova:2021fnr}. Recall that in an IQFT the two-body scattering matrix fully characterises all scattering processes of the theory and can be almost entirely determined by a consistency procedure known as \emph{bootstrap} \cite{Zamolodchikov:1978xm, Zamolodchikov:1989hfa}. The ``almost'' here refers to the so-called \emph{CDD ambiguity} \cite{Zamolodchikov:1978xm}: the boostrap equations admit a ``minimal solution'' (i.e., a solution having the minimal set of singularities) which can then be dressed by an arbitrary number of CDD factors \cite{Castillejo:1955ed}. These are trivial solutions of the $S$-matrix bootstrap equations, meaning that although they can modify the $S$-matrix in non-trivial ways, they do not include poles in the physical strip, thus leaving the spectrum of the theory intact. Let us now make all these ideas more precise. 

\medskip 

For simplicity, we are going to focus IQFTs with single particle spectrum. After a $\mathrm{T}\overline{\mathrm{T}}$ perturbation the scattering matrix $S_0(\theta)$, where $\theta$ is the rapidity variable, is modified to
\beq
S_{\alpha}(\theta):= S_0(\theta) e^{-i\alpha m^2 \sinh\theta}\,.
\label{1}
\eeq 
In the case of a generalised $\mathrm{T}\overline{\mathrm{T}}$ deformation, in the sense defined above, the scattering matrix will be dressed by a generic CDD factor $\Phi_{\bal}(\theta)$
\beq
S_{\boldsymbol{\alpha}}(\theta) := S_{\boldsymbol{0}}(\theta) \Phi_{\boldsymbol{\alpha}}(\theta)\, \quad {\mathrm{with}} \quad \log \Phi_{\boldsymbol{\alpha}}(\theta)=-i\sum\limits_{s\in \mathcal{S}}\alpha_s m^{2s} \sinh(s\theta)\,.
\label{2}
\eeq 
Here $\mathcal{S}$ denotes the set of the spins of the local conserved quantities and it is a fundamental \emph{datum} that depends on the specific theory under consideration. In the interpretation of IQFTs as massive perturbations of CFTs, these are the conserved charges that are not destroyed by the deformation process \cite{Zamolodchikov:1989hfa, Negro:2016yuu}. In many cases, such as the thermal Ising and sinh-Gordon models, $\mathcal{S}$ coincides with the set of odd natural numbers, but this is not always the case\footnote{For example in the magnetic deformation of Ising model, $\mathcal{S} = \{1,7,11,13,17,19,23,29\}\,\mathrm{mod}\,30$ \cite{Zamolodchikov:1989hfa}. Remarkably, these numbers are the \emph{Coxeter exponents} of the Lie algebra $E_8$. This is just one instance of the deep relationship between IQFTs and the structure of Lie algebras, particularly evident for models in the Toda family. The interested reader can find more details in, e.g., \cite{Christe:1989my, Braden:1989bu,FringKorff}. }. In \eqref{2}, $\boldsymbol{\alpha}$ is a short-hand for the set $\{\alpha_s\, \vert\, s \in \mathcal{S}\}$ and in \eqref{1} $\alpha=\alpha_1$. The $\alpha_s$ are coupling constants such that $\alpha_s m^{2s}$ is adimensional, with $m$ being a fundamental mass scale. Written in terms of energy-momentum vectors of the two scattering particles $p_{\mu}^{i} = m(\cosh\theta_i, \sinh\theta_i)$, with $i=1,2$, the $\mathrm{T}\overline{\mathrm{T}}$ deformation \eqref{1} reads $\exp[-i\alpha \varepsilon^{\mu\nu} p_\mu^1p_\nu^2]$, where $\theta=\theta_1-\theta_2$. For simplicity, throughout this paper we will take the mass scale $m=1$.

The CDD factors $\Phi_{\boldsymbol{\alpha}}$ in \eqref{2} automatically satisfy all $S$-matrix consistency equations, i.e. unitarity and crossing. However they introduce a very uncommon double-exponential dependence on the rapidity that radically changes the $S$-matrix asymptotic behaviour and has a stark effect on the theory's RG flow. Generally speaking, whereas  the IR regime is left unaltered by the perturbation, in the short-distance limit the theory displays unusual features, which are incompatible with the existence of a UV CFT \cite{Dubovsky:2012wk, Dubovsky:2013ira, Dubovsky:2017cnj}. This is in agreement with the fact that the CDD factors \eqref{2} correspond to perturbations by irrelevant operators, whose presence is not felt at large distances but is expected to severely alter the properties of the UV. Generalised $\mathrm{T}\overline{\mathrm{T}}$ deformations and their properties have been studied from several viewpoints: in the context of 2D classical and quantum field theory\footnote{While extensions of the $\mathrm{T}\overline{\mathrm{T}}$ deformation to $1$-dimensional, quantum-mechanical systems yields well-defined, controllable theories \cite{Gross:2019ach, Gross:2019uxi, ste_zeta}, the proposed higher dimensional generalisations \cite{Bonelli:2018kik, Taylor:2018xcy} present several complications, mainly consequence of the fewer amount of restrictions on the short-distance singularities arising in the OPE of energy-momentum tensor components. Some promising advances have been made in \cite{Conti:2022egv} that considers extensions of the $\mathrm{T}\overline{\mathrm{T}}$ deformation in higher dimensions as field-dependent perturbations of the space-time geometry, in the same spirit as the geometric interpretation of \cite{Conti:2018tca} and the topological gravity picture of \cite{Dubovsky:2018bmo}} \cite{Conti:2018jho, Conti:2018tca, Dubovsky:2023lza}, in the framework of the ODE/IM correspondence \cite{Aramini:2022wbn} (see \cite{Dorey:2019ngq} for a review), employing the thermodynamic Bethe ansatz (TBA) approach\footnote{Remarkably, the first TBA analysis on models with S-matrices of the type \eqref{1} were performed in \cite{Dubovsky:2012wk, Caselle:2013dra}, before the ``official'' introduction of the $\mathrm{T}\overline{\mathrm{T}}$ deformation in \cite{Smirnov:2016lqw, Cavaglia:2016oda}.} \cite{Cavaglia:2016oda, Conti:2019dxg, Hernandez-Chifflet:2019sua, Camilo:2021gro, Cordova:2021fnr, LeClair:2021opx, LeClair:2021wfd, Ahn:2022pia}, from the viewpoint of perturbed conformal field theory \cite{Guica:2017lia, Cardy:2018sdv, Cardy:2019qao, Aharony:2018vux, Aharony:2018bad, Guica:2020uhm, Guica:2021pzy, Guica:2022gts} and also in the context of string theory \cite{Baggio:2018gct, Dei:2018jyj, Chakraborty:2019mdf, Callebaut:2019omt}, holography \cite{McGough:2016lol, Giveon:2017nie, Gorbenko:2018oov, Kraus:2018xrn, Hartman:2018tkw, Guica:2019nzm, Jiang:2019tcq,Jafari:2019qns}, quantum gravity \cite{Dubovsky:2017cnj, Dubovsky:2018bmo, Tolley:2019nmm, Iliesiu:2020zld, Okumura:2020dzb, Ebert:2022ehb}, out-of-equilibrium conformal field theory \cite{Medenjak:2020ppv, Medenjak:2020bpe}, long-range spin chains \cite{Bargheer:2008jt,Bargheer:2009xy,PJG, Marchetto:2019yyt}, and the generalised hydrodynamics (GHD) approach \cite{Cardy:2020olv}.

In an IQFT several important physical observables can be computed exactly using the TBA approach \cite{tba1, tba2}. In particular, a key quantity is the ground state energy of the theory on a cylinder of radius $R$, denoted by $E(R)$. In a conventional, UV-complete IQFT the small $R$ behaviour of $E(R)$ is dominated by a simple pole $E(R)\sim -\frac{\pi c}{6R}$ that signals the presence of a CFT with central charge $c$ in the UV. It has been shown \cite{Cavaglia:2016oda} that in $\mathrm{T}\overline{\mathrm{T}}$-deformed theories there are two possible scenarios for small $R$ properties of $E(R,\alpha)$, depending to the sign of $\alpha$ in \eqref{1}:
\begin{itemize}
    \item for $\alpha > 0$ the UV limit of the ground-state energy is finite $\lim_{R\rightarrow 0} E(R,\alpha) = -e_0(\alpha)$, implying that the theory possesses a finite amount of degrees of freedom;
    \item for $\alpha < 0$ a square-root branch point appears at  $R=R_{\ast}\sim \vert\alpha\vert^{-1/2}>0$, signalling the presence of a Hagedorn growth \cite{Hagedorn:1965st} of the density of states at high energy.
\end{itemize}
The GHD viewpoint \cite{Cardy:2020olv} complements this picture, suggesting that the abnormal UV limit of $\mathrm{T}\overline{\mathrm{T}}$-deformed IQFTs can be imputed to point-like particles acquiring a positive or negative size, for $\alpha>0$ and $\alpha<0$ respectively. For what concerns the generalised $\mathrm{T}\overline{\mathrm{T}}$ deformations with $S$-matrices given by \eqref{2}, it was shown \cite{Camilo:2021gro} that they can only display the second of the above behaviours. In fact, for $S$-matrices \eqref{2} with a finite set of couplings $\bal$, the standard TBA equations only make sense if the coupling with larger index $s$ is negative $\alpha_s<0$. One possible way to study the regime $\alpha_s>0$ is to employ the \emph{generalised TBA} \cite{Mossel:2012vp} approach. Note that neither of the behaviours listed above is compatible with Wilson’s paradigm of local QFTs \cite{Wilson:1973jj}. For this reason -- and thanks to their property of being solvable and controllable at any energy scale -- the generalised $\mathrm{T}\overline{\mathrm{T}}$-deformed theories can be considered as a sensible extension of the Wilsonian notion of a local QFT.
\medskip

In this paper we embark on the study of generalised $\mathrm{T}\overline{\mathrm{T}}$-perturbed IQFTs by employing a traditional approach in the IQFT context, namely the form factor program for matrix elements of local fields. Our work constitutes the first systematic attempt (in the sense that it encompasses a large set of theories, fields and perturbations) at computing the form factors of local and semi-local fields in generalised $\mathrm{T}\overline{\mathrm{T}}$-perturbed IQFTs. Although the present paper focuses particularly on the Ising model some of the key results are clearly more general, as discussed in \cite{PRL}. 

The $n$-particle form factor of a field $\mathcal{O}$ is defined as:
\beq 
F_n^{\mathcal{O}}(\theta_1,\ldots,\theta_n;\boldsymbol{\alpha}):=
\bra 0| \mathcal{O}(0)|\theta_1,\ldots,\theta_n|0\ket\,,
\label{FFs}
\eeq 
where $|\theta_1,\ldots,\theta_n|0\ket$ is a state of of $n$ incoming particles of rapidites $\{\theta_i\}_n$ and $|0\ket$ is the vacuum state. Assuming that the form factor equations still hold for the fields of the perturbed theory, the quantities \eqref{FFs} can be computed by following the usual bootstrap program \cite{Karowski:1978vz, smirnov1992form}, taking as starting point the deformed $S$-matrix \eqref{2}. They can then be used as building blocks to construct correlation functions, allowing in particular to study the short and long-distance limits of these correlation functions and to understand how those limits are affected by the perturbation. The form factors and correlation functions of a theory with similar unusual features (the sinh-Gordon model, with deformation $\Phi_{\bal}(\theta)=-1$) were computed for the first time in \cite{sGMuss}, and found to give rise to a number of problems, including non-covergent correlators and form factors involving free parameters that could not be fully fixed. The present work will allow us to understand similar issues in a much wider context.

%\medskip
Given the unusual properties of $\mathrm{T}\overline{\mathrm{T}}$-perturbed IQFTs, notably the lack of a proper UV fixed point, it is natural to ask whether a form factor program can be pursued in the first place. We believe that the results of this work represent an affirmative answer to this question as well as the beginning of a new research program. Indeed, as discussed earlier, there has already been a vast amount of work on the description and interpretation of the scattering and thermodynamic properties of the $S$-matrices \eqref{2}, work that has been possible despite the unusual properties of the resulting models. With the first step of the bootstrap program now completed, it is then natural to proceed to the next stage of the program, namely the computation of form factors and correlation functions. This further step will provide new insights into the physics of the $\TTb$ perturbations and of its generalisations.

\medskip

This paper is organised as follows: In Section 2 we review the form factor program for IQFTs, extending it to generalised $\mathrm{T}\overline{\mathrm{T}}$-perturbed IQFTs. In particular, we find a general formula for the minimal form factor in any IQFT with diagonal scattering. In Section 3 we write the form factor equations for higher particle numbers and general local and semi-local fields and $S$-matrix, assuming a single-particle spectrum, for simplicity. We then specialise to the Ising field theory and find closed solutions to the deformed form factor equations for the field $\Theta$ (the trace of the stress-energy tensor) and the fields $\mu$ and $\sigma$, typically known as the disorder and order field, respectively. 
In Section 4 we explain how our form factor solutions can be used as building blocks for correlation functions and we analyse their asymptotic properties both at short and long distances. We also discuss the convergence -- or the lack thereof, depending on the sign of the deformation parameters -- of the form factor expansion and show how for a certain range of values, the form factor series is convergent. In such cases traditional consistency checks of the form factor solutions, such as the $\Delta$-sum rule and Zamolodchikov's $c$-theorem, are put to the test and the results discussed. 
We conclude in Section 5. The Appendix contains an extension of the discussion in Section 4. 

\section{Form Factor Program for  Generalised $\mathrm{T}\overline{\mathrm{T}}$-Deformed Theories}
Consider an IQFT with a single-particle spectrum and deformed $S$-matrix given by \eqref{2}, with $m=1$. In the absence of bound states, the form factor equations for local and semi-local fields in this theory can be written as \cite{Karowski:1978vz, smirnov1992form}:
\beq 
F_n^{\mathcal{O}}(\theta_1,\ldots,\theta_i,\theta_{i+1},\ldots,\theta_n;\boldsymbol{\alpha})=S_{\bal}(\theta_{i}-\theta_{i+1}) F_n^{\mathcal{O}}(\theta_1,\ldots,\theta_{i+1},\theta_{i}, \ldots, \theta_n;\boldsymbol{\alpha})\,,
\label{W1}
\eeq 
\beq 
F_n^{\mathcal{O}}(\theta_1+2\pi i , \theta_2 \ldots,\theta_n;\boldsymbol{\alpha})= \gamma_\mathcal{O} F_n^{\mathcal{O}}(\theta_2, \ldots, \theta_n,\theta_1;\boldsymbol{\alpha})\,,
\label{W2}
\eeq 
and 
\beq
 \lim_{\bar{\theta}\rightarrow \theta} (\bar{\theta}-\theta) F^{\mathcal{O}}_{n+2}(\bar{\theta}+i\pi, \theta, \theta_1,\ldots, \theta_n;\boldsymbol{\alpha}) = i \left(1-\gamma_{\mathcal{O}}\prod_{j=1}^n S_\alpha(\theta-\theta_j)\right) F^{\mathcal{O}}_n(\theta_1,\ldots, \theta_n;\boldsymbol{\alpha})\,.
 \label{KRE}
 \eeq
The two first equations constrain the monodromy of the form factors while the last one, the kinematic residue equation, specifies their pole structure. In \eqref{KRE} we introduced the parameter $\gamma_{\mathcal{O}}$, a complex number of unit length $\left\vert\gamma_{\mathcal{O}}\right\vert = 1$, known as the factor of local commutativity in \cite{Yurov:1990kv}. In theories possessing an internal symmetry, such as the Ising model, it can be a non-trivial phase.

\medskip 

The solution procedure typically starts with finding a ``minimal" solution to the two-particle form factor equations\footnote{For fields that are primary in the conformal limit and their descendants, the zero-particle form factor is the vacuum expectation value of the field $\bra \mathcal{O}\ket$. For spinless fields, the one-particle form factor is also a constant. Thus, the two-particle form factor is the first non-trivial solution to the equations.}
\beq
F_{\rm{min}}(\theta;\boldsymbol{\alpha})=S_{\boldsymbol{\alpha}}(\theta) F_{\rm{min}}(-\theta;\boldsymbol{\alpha})=F_{\rm{min}}(2\pi i -\theta;\boldsymbol{\alpha})\,,
\eeq 
that presents no poles in the physical strip. Since the $S$-matrix is factorised as in \eqref{2}, it is natural to make the ansatz
\beq 
F_{\rm{min}}(\theta;\boldsymbol{\alpha}):= F_{\rm{min}}(\theta;\boldsymbol{0}) \varphi(\theta;\boldsymbol{\alpha})\,,
\label{fact}
\eeq 
where the function $\varphi(\theta; \boldsymbol{\alpha})$ solves the equations
\beq
\varphi(\theta;\boldsymbol{\alpha})=\Phi_{\boldsymbol{\alpha}}(\theta) \varphi(-\theta;\boldsymbol{\alpha})=\varphi(2\pi i -\theta;\boldsymbol{\alpha})\,.
\label{varphi}
\eeq 
In fact, $\varphi(\theta;\boldsymbol{\alpha})$ is the minimal form factor of the generalised $\mathrm{T}\overline{\mathrm{T}}$-deformed free boson theory. It is easy to show that the simplest solution to \eqref{varphi} is
\beq 
\varphi(\theta;\boldsymbol{\alpha})=\exp\left[-{\frac{ (i\pi-\theta)}{2\pi} \sum_{s\in \mathcal{S}} \alpha_s \sinh (s\theta)}\right]\,.\label{varphi2}
\eeq
More generally, given a solution \eqref{varphi2} it is always possible to multiply it by a function of the following form
\beq 
C(\theta;\bel):=\exp\left[\sum_{s\in \mathbb{Z}} \beta_s \cosh(s\theta) \right]\,,
\label{betas}
\eeq 
which plays the role of a ``CDD factor" for the minimal form factor itself. Once \eqref{varphi2} is known, the minimal form factor of any theory with a single-particle spectrum can be easily constructed through \eqref{fact} and the generalisation to multiple particle types follows naturally. In this paper, we will not consider the larger family of solutions generated by \eqref{betas}, however it is possible to give a physical interpretation for the factor $C(\theta;\bel)$. The terms 
$ \alpha_s \sinh (s\theta)$ in the sum over $s$ are the one-particle eingenvalues of local conserved quantities of spin $s$. However, those one-particle eigenvalues are more generically linear combinations of the form $\alpha_s \sinh(s\theta)+ \beta_s \cosh(s\theta)$. The fact that we can always add a sum of terms $\beta_s \cosh (s\theta)$ is a reflection of this ``ambiguity" in our choice of the conserved quantities. It also means that given a solution to the form factor equations for a certain field for the choice $\bel=\bol$, new solutions can be constructed by ``switching on" some of the parameters $\bel$ (although in some cases, depending on which values $s$ can take in $\beta_s$ the introduction of parameters $\beta_s$ can change not just the minimal form factor but other parts of the form factor too). These new solutions, depending on different choices of the parameters $\bel$, are distinct solutions to the same form factor equations, thus will correspond to a different field. Interestingly all of these fields will ``flow" to the original unperturbed field $\mathcal{O}$ when the parameters $\bal, \bel \mapsto \bol$. 

\medskip
For the free boson ($+$) and free fermion ($-$) theories, with scattering matrices $S_{\boldsymbol{0}}(\theta)=\pm 1$, the minimal form factors are
\beqa
F_{\rm{min}}^+(\theta;\boldsymbol{\alpha})=\varphi(\theta;\boldsymbol{\alpha})\,, \qquad
F_{\rm{min}}^-(\theta;{\bf \alpha})=-i\sinh\frac{\theta}{2}\varphi(\theta;\boldsymbol{\alpha})\,,
\label{fminus}
\eeqa 
whereas for an interacting theory, such as the sinh-Gordon model with scattering matrix
\beq
S_{\boldsymbol{0}}(\theta)=\frac{\tanh\frac{1}{2}\left(\theta-\frac{i\pi B}{2}\right)}{\tanh\frac{1}{2}\left(\theta+\frac{i\pi B}{2}\right)}\qquad {\mathrm{with}} \qquad B\in [0,2]\,,
\eeq 
the minimal form factor is
\beq
F_{\rm{min}}^{\rm sG}(\theta;\boldsymbol{\alpha})=F^{\rm sG}_{\rm{min}}(\theta;\boldsymbol{0})\varphi(\theta;\boldsymbol{\alpha})\, \label{sG}
\eeq 
where the underformed form factor $F^{\rm sG}_{\rm{min}}(\theta;\boldsymbol{0})$ is well-known \cite{FMS, KK, freefield1}.  The key observation is then that the function $\varphi(\theta;\bal)$ is common to all theories and describes the universal way in which the minimal solution to the form factor equations is modified by the presence of irrelevant perturbations. We are now in a position where we can systematically study any IQFT with diagonal scattering. 

Before moving on, we remark that, if $\mathcal{S}$ contains only odd spins, as we are going to assume henceforth, the function $\varphi$ satisfies the following identity that will be useful in subsequent computations:
\beqa
\varphi(\theta;\boldsymbol{\alpha})\varphi(\theta+i\pi;\boldsymbol{\alpha})=\sqrt{\Phi_{\boldsymbol{\alpha}}(\theta)}\,,
\label{prod}
\eeqa 
Furthermore, with our normalisations 
\beq
\varphi(i\pi;{\boldsymbol{\alpha}})=\varphi(\theta;\bol)=1\,.
\label{fmin1}
\eeq
We now turn to the problem of computing higher particle form factors.

\section{Solving the Recursive Equations}
Consider the form factor equations \eqref{W1}, \eqref{W2} and \eqref{KRE} for higher particle form factors of local or semi-local operators. Assuming there is a kinematic pole whenever any two rapidities differ by $i\pi$\footnote{Note that this is typically not the case for the two-particle form factor of the trace of the stress-energy tensor $\Theta$ or more generally of any local fields (i.e. fields where $\gamma_{\mathcal{O}}=1$), since in this case the r.h.s. of the kinematic residue equation \eqref{KRE} with $n=2$ is vanishing. The kinematic pole in the two-particle form factor is generally only present for symmetry fields, which sit at the origin of branch cuts in the complex plane. However, for theories with a non-trivial $S$-matrix, kinematic residue poles will be present for the higher particle form factors, even for local fields.}, we can make the standard type of ansatz
\beq
F^{\mathcal{O}}_n(\theta_1,\ldots,\theta_n; \boldsymbol{\alpha})=H^{\mathcal{O}}_n Q^{\mathcal{O}}_n(x_1,\ldots,x_n;\boldsymbol{\alpha}) \prod_{i<j} \frac{F_{\rm min}(\theta_{ij};\boldsymbol{\alpha})}{x_i+x_j}\,.\label{ansatz}
\eeq
The functions $Q_n^\mathcal{O}(x_1,\ldots,x_n;\boldsymbol{\alpha})$ are symmetric and $2\pi i$-periodic functions of all the variables $x_j:=e^{\theta_j}$.  Since they are symmetric, they can be written in terms of a basis of elementary symmetric polynomials in these variables. Denoting the order $j$ elementary symmetric polynomial of $n$ variables as $\sigma_j^{(n)}(x_1,\ldots, x_n)$, their definition is as follows
\beqa
\sigma_0^{(n)}(x_1,\ldots, x_n)=1 \qquad \mathrm{and} \qquad  
 \sigma_j^{(n)}(x_1,\ldots, x_n)=\sum_{1\leq i_1<i_2<\cdots <i_j \leq n} x_{i_1} x_{i_2}\cdots x_{i_j}\,.
 \label{elementary}
 \eeqa
The symbols $H_n^\mathcal{O}$ in \eqref{ansatz} are constants, independent of the rapidity variables. 
Plugging this ansatz into the kinematic residue equation \eqref{KRE}
we can reshape the left hand side into the following form
\beqa
&& F_{n+2}^{\mathcal{O}}(\bar{\theta}+i\pi, \theta, \theta_1,\ldots, \theta_n;\boldsymbol{\alpha}) = H^{\mathcal{O}}_{n+2} Q^{\mathcal{O}}_{n+2}(-\bar{x},x,x_1,\ldots,x_n;\boldsymbol{\alpha}) \prod_{1\leq i<j \leq n}\frac{F_{\rm min}(\theta_{ij};\boldsymbol{\alpha})}{x_i+x_j} \nonumber\\
&& \qquad \qquad \times  \left[\prod_{j=1}^n \frac{F_{\rm min}(\theta-\theta_j;\boldsymbol{\alpha})F_{\rm min}(\bar{\theta}+i\pi-\theta_j;\boldsymbol{\alpha})}{(x+x_j)(-\bar{x}+x_j)}\right]\frac{F_{\rm min}(i\pi;\boldsymbol{\alpha})}{-\bar{x}+x}\,
 \eeqa
 where $x=e^\theta$ and $\theta_{ij}:=\theta_i-\theta_j$. Then the equation becomes
  \beqa 
 &&\!\!\!\! F_{\rm min}(i\pi;\boldsymbol{\alpha}) \lim_{\bar{\theta}\rightarrow \theta} \frac{\bar{\theta}-\theta}{x-\bar{x}} H^{\mathcal{O}}_{n+2} Q^{\mathcal{O}}_{n+2}(-\bar{x},x,x_1,\ldots,x_n;\boldsymbol{\alpha}) \prod_{j=1}^n \frac{F_{\rm min}(\theta-\theta_j;\boldsymbol{\alpha})F_{\rm min}(\bar{\theta}+i\pi-\theta_j;\boldsymbol{\alpha})}{(x+x_j)(-\bar{x}+x_j)} \nonumber\\
&& = i \left(1-\gamma_{\mathcal{O}}\prod_{j=1}^n S_{\boldsymbol{\alpha}}(\theta-\theta_j)\right)  H^{\mathcal{O}}_n Q^{\mathcal{O}}_n (x_1,\ldots, x_n;\boldsymbol{\alpha}).
\label{KRE2}
 \eeqa
 where we used \eqref{fmin1}.
Finally, noting that 
$
 \lim_{\bar{\theta}\rightarrow \theta} \frac{\bar{\theta}-\theta}{x-\bar{x}}=-\frac{1}{x}$ and using the properties \eqref{fact}, \eqref{prod} and \eqref{fmin1}, we arrive at
 \beqa 
 && -x^{-1} F_{\rm min}(i\pi;\boldsymbol{\alpha}) H^{\mathcal{O}}_{n+2} Q^{\mathcal{O}}_{n+2}(-{x},x,x_1,\ldots,x_n;\boldsymbol{\alpha}) \prod_{j=1}^n \frac{F_{\rm min}(\theta-\theta_j;\boldsymbol{0})F_{\rm min}(\bar{\theta}+i\pi-\theta_j;\boldsymbol{0})}{x_j^2-x^2} \nonumber\\
&& = i \left[\prod_{j=1}^n \Phi_{\boldsymbol{\alpha}}(\theta-\theta_j)^{-\frac{1}{2}} -\gamma_{\mathcal{O}}\prod_{j=1}^n \Phi_{\boldsymbol{\alpha}}(\theta-\theta_j)^{\frac{1}{2}}S_{\boldsymbol{0}}(\theta-\theta_j)\right]  H^{\mathcal{O}}_n Q^{\mathcal{O}}_n (x_1,\ldots, x_n;\boldsymbol{\alpha}).
\label{KRE3}
 \eeqa
 We will now show how to find solutions to this equation for the Ising field theory.
 
\subsection{The Ising Field Theory}
The form factors of the Ising field theory have been extensively studied in the literature \cite{ZI, ST}. In the so-called disordered phase, form factors and correlation functions of the order ($\sigma$) and disorder ($\mu$) fields were computed in \cite{Yurov:1990kv, BB}. Around the same time, form factors of descendant fields in the CFT sense where computed in \cite{cardymussardo} and shown to be in one-to-one correspondence with the Virasoro irreducible representations characterising the critical theory. More recently, the study of form factors has expanded to encompass other types of fields which are present in the replica version of the Ising model. Form factors of the branch point twist field where computed \cite{entropy, nexttonext} as well as for composite branch point twist fields in \cite{SymResFF}. These form factors play a prominent role in the study of entanglement measures of IQFT. 

The Ising field theory is a free fermion theory with scattering matrix $S_{\boldsymbol{0}}(\theta)=-1$, and $F_{\rm min}(\theta;\boldsymbol{0})=-i\sinh\frac{\theta}{2}$. For this model, the set $\mathcal{S}$ of the spins of local conserved quantities coincides with the odd positive integers
\beq 
\mathcal{S}=\{s=2n-1| n\in \mathbb{Z}^+\}\,,
\eeq 
With these specifications, equation \eqref{KRE3} simplifies to
 \beqa 
 && -x^{-1} H^{\mathcal{O}}_{n+2} Q^{\mathcal{O}}_{n+2}(-{x},x,x_1,\ldots,x_n;\boldsymbol{\alpha}) \prod_{j=1}^n \frac{i}{4 x x_j} \nonumber\\
&& = i \left[\prod_{j=1}^n \Phi_{\boldsymbol{\alpha}}(\theta-\theta_j)^{-\frac{1}{2}} -\gamma_{\mathcal{O}} (-1)^n \prod_{j=1}^n \Phi_{\boldsymbol{\alpha}}(\theta-\theta_j)^{\frac{1}{2}}\right]  H^{\mathcal{O}}_n Q^{\mathcal{O}}_n (x_1,\ldots, x_n;\boldsymbol{\alpha}).
\label{KRE4}
 \eeqa
that we can further split into two equations:
\beqa 
 && Q^{\mathcal{O}}_{n+2}(-{x},x,x_1,\ldots,x_n;\boldsymbol{\alpha}) =  x^{n+1} \prod_{j=1}^n x_j  \nonumber\\
&& \times \left[\prod_{j=1}^n \Phi_{\boldsymbol{\alpha}}(\theta-\theta_j)^{-\frac{1}{2}} -\gamma_{\mathcal{O}} (-1)^n \prod_{j=1}^n \Phi_{\boldsymbol{\alpha}}(\theta-\theta_j)^{\frac{1}{2}}\right]   Q^{\mathcal{O}}_n (x_1,\ldots, x_n;\boldsymbol{\alpha}).
\label{QQ}
 \eeqa
 and 
 \beq 
 H^{\mathcal{O}}_{n+2} =  4^n i^{-1-n} H^{\mathcal{O}}_{n}\,.
 \label{eq:Hconst_eq}
 \eeq 
 The latter is easily solved
 \beq 
  H^{\mathcal{O}}_{2n}= 4^{n(n-1)} i^{-n^2} H^{\mathcal{O}}_{0}\quad \mathrm{and} \quad H^{\mathcal{O}}_{2n+1}=4^{n^2} i^{-n(n+1)} H^{\mathcal{O}}_{1}\,.
  \label{HH}
 \eeq
whereas the equation for the polynomials 
$Q^{\mathcal{O}}_{n}(x_1,\ldots,x_n;\boldsymbol{\alpha})$ requires some additional input about the field $\mathcal{O}$. Let us recall the field content of the Ising field theory (besides the fermion itself) consists of the energy $\varepsilon$, the spin $\sigma$ and the disorder $\mu$ fields. Their form factors are well known
\beq 
F^\mu_{2n}(\theta_1,\ldots, \theta_{2n};\boldsymbol{0})=i^n \prod_{i<j}\tanh\frac{\theta_{ij}}{2}\,, \quad F^\mu_{2n+1}(\theta_1,\ldots, \theta_{2n+1};\boldsymbol{0})=0\,,
\label{mu}
\eeq 
\beq 
F^\sigma_{2n}(\theta_1,\ldots, \theta_{2n};\boldsymbol{0})=0\,, \quad F^\sigma_{2n+1}(\theta_1,\ldots, \theta_{2n+1};\boldsymbol{0})=i^n \prod_{i<j}\tanh\frac{\theta_{ij}}{2}\,,
\label{sigma}
\eeq 
and 
\beq 
F^\varepsilon_2(\theta;\boldsymbol{0})=-i \sinh\frac{\theta}{2}\,, \quad F^\varepsilon_{n>2}(\theta_1,\ldots, \theta_{n};\boldsymbol{0})=0\,.
\eeq 
In the Ising model, there is a $\mathbb{Z}_2$ symmetry and the fields organise themselves into two symmetry sectors. The energy $\varepsilon$  and the spin $\sigma$ fields both have $\gamma_\varepsilon=\gamma_\sigma=1$ whereas the field $\mu$ has $\gamma_\mu=-1$. Due to the symmetry the fields $\varepsilon$ and $\mu$ have only non-vanishing even-particle form factors (more precisely, since $\varepsilon$ is a bilinear in the fermion field it only has non-vanishing two-particle form factor) whereas $\sigma$ has only odd-particle ones. The generalised $\mathrm{T}\overline{\mathrm{T}}$ deformations do not break the $\mathbb{Z}_2$ symmetry so these properties are all preserved. Instead of focusing on the energy field itself, it is common to study the properties of the trace of the stress-energy tensor $\Theta$ which is a field proportional to $\varepsilon$. Its form factors are very simple
\beq 
F^\Theta_2(\theta;\boldsymbol{0})=-2\pi i  \sinh\frac{\theta}{2}\,, \quad F^\Theta_{n>2}(\theta_1,\ldots, \theta_{n};\boldsymbol{0})=0\,.
\label{FT0}
\eeq 
For dimensional reasons, the form factor above must be proportional to $m^2$ (recall that we have set $m=1$ throughout the paper). Consequently, in the massless limit $m\rightarrow 0$ all the form factors of $\Theta$ vanish identically, which is consistent with vanishing of the stress-energy tensor trace in the CFT limit. 

For $\Theta$ the equation \eqref{QQ} becomes
\beqa
&& Q^{\Theta}_{2n+2}(-{x},x,x_1,\ldots,x_{2n};\boldsymbol{\alpha})\nonumber\\
&&  2i x^{2n+1} \sin\left(\sum_{s \in \mathcal{S}} \frac{\alpha_s}{2} \left(\sum_{j=1}^{2n} \sinh s(\theta-\theta_j)\right)\right) \prod_{j=1}^{2n} (x_j) Q^{\Theta}_{2n} (x_1,\ldots, x_{2n};\boldsymbol{\alpha})\label{thefieldE}
\eeqa 
and
 \beq 
 H^{\Theta}_{2n}=4^{n(n-1)} i^{-n^2} H_0^\Theta \quad \mathrm{with} \quad H_0^\Theta=F_0^\Theta\,.
 \eeq 
 The equation \eqref{thefieldE} only holds for $n>0$. 
 Indeed, the operator $\Theta$ is special in that its two-particle form factor has no kinematic pole while higher particle ones do. Hence, we expect to have the standard normalisation $F_0^\Theta=2\pi=F_2^\Theta(i\pi;\bal)$. The two-particle form factor should be generalisation of \eqref{FT0} with the standard form
 \beq 
F^\Theta_2(\theta;\boldsymbol{\alpha})=2\pi  f(\theta;\bal) \frac{F_{\rm min}(\theta;\boldsymbol{\alpha})}{F_{\rm min}(i\pi;\boldsymbol{\alpha})}\,.
\label{FTa}
\eeq 
where the function $f(\theta;\bal)$ is constrained by $f(i\pi;\bal)=f(\theta;\bol)=1$ and should be an even, $2\pi i$-periodic function of $\theta$. We will see later that consistency with higher particle solutions requires this function to be non-trivial. The equation \eqref{thefieldE} determines the higher-particle form factors which, as the sine factor suggests, will all vanish for $\alpha_i=0$, in agreement with \eqref{FT0}. 

For the field $\mu$ we have $\gamma_\mu=-1$ and only even-particle form factors. Then the equations (\ref{QQ}, \ref{HH}) become:
\beqa
&& Q^\mu_{2n+2}(-{x},x,x_1,\ldots,x_{2n};\boldsymbol{\alpha})=\nonumber\\
&& 2 x^{2n+1} \cos\left(\sum_{s \in \mathcal{S}} \frac{\alpha_s}{2} \left(\sum_{j=1}^{2n} \sinh s(\theta-\theta_j)\right)\right)  \prod_{j=1}^{2n} (x_j) Q^\mu_{2n} (x_1,\ldots, x_{2n};\boldsymbol{\alpha})\,,
\label{thefieldmu}
\eeqa 
and 
 \beq 
 H^{\mu}_{2n}=4^{n(n-1)} i^{-n^2} H_0^\mu \quad \mathrm{with} \quad H_0^\mu=\bra \mu \ket_{\bal}\,.
 \eeq 
 where $\bra \mu \ket_{\bal}=F_0^\mu(\bal)$ is the vacuum expectation value of $\mu$, which may depend on $\bal$ in the deformed theory. 
In the case of the field $\sigma$, we only have odd particle numbers and a trivial factor of local commutativity $\gamma_\sigma=1$, so we also get a cosine
\beqa
&& Q^\sigma_{2n+1}(-{x},x,x_1,\ldots,x_{2n+1};\boldsymbol{\alpha})= \nonumber\\ 
&& 2 x^{2n} \cos\left(\sum_{s \in \mathcal{S}} \frac{\alpha_s}{2} \left(\sum_{j=1}^{2n-1} \sinh s(\theta-\theta_j)\right)\right)   \prod_{j=1}^{2n-1} (x_j) Q^\sigma_{2n-1} (x_1,\ldots, x_{2n-1};\boldsymbol{\alpha})\nonumber
\eeqa 
and
\beq 
 H^{\sigma}_{2n+1}=4^{(n-1)^2} i^{-n(n-1)}   H_1^\sigma \quad \mathrm{with} \quad H_1^\sigma=F_1^\sigma(\bal)\,.
 \eeq 

\subsubsection{Solving the Equations: The Fields $\mu$ and $\sigma$}

As shown in \cite{PRL}, solutions to the form factor equations above factorise into the unperturbed solutions times a function that depends on the scattering matrix and the locality properties of the field. This factorisation is particularly natural for the Ising field theory, as we will see below.  We can write
\beq
Q_{2n}^\mu(x_1,\ldots,x_{2n};\bal):= D_{2n}^\mu(x_1,\ldots,x_{2n};\bal) P_{2n}^\mu(x_1,\ldots,x_{2n})\,,
\label{Factmu}
\eeq
where
\beq
D_{2n+2}^\mu(-x,x,x_1,\ldots,x_{2n};\bal)=2\cos\left(\sum_{s \in \mathcal{S}} \frac{\alpha_s}{2} \left(\sum_{j=1}^{2n} \sinh s(\theta-\theta_j)\right)\right) D_{2n}^\mu(x_1,\ldots,x_{2n};\bal)\,,
\label{Dmu}
\eeq 
and
\beqa
P_{2n+2}^\mu(-{x},x,x_1,\ldots,x_{2n})=
 x^{2n+1} \sigma_{2n}^{(2n)} P^\mu_{2n} (x_1,\ldots, x_{2n})\,.\label{eqP}
\eeqa 
The equations for the operator $\sigma$ are identical to (\ref{Factmu} -- \ref{eqP}) with $2n \mapsto 2n-1$ and so are the solutions. The equation \eqref{eqP} can be solved easily, starting with the cases
\beq 
P_2^\mu(-x,x)=x P_0^\mu \quad \mathrm{and} \quad P_3^\sigma(-x,x,x_1)=x^2 x_1 P_1^\sigma\,,
\eeq 
and proceeding recursively. For instance
\beq 
P^\mu_2(x_1,x_2)=-i(\sigma_2^{(2)})^{\frac{1}{2}} P_0^\mu\,,
\eeq 
and so
\beq 
P_4^\mu(-x,x,x_1,x_2)= -i(x^2 x_1 x_2)^{\frac{3}{2}}P_0^\mu\quad \Rightarrow \quad
P_4^\mu(x_1,x_2,x_3,x_4)=(\sigma_4^{(4)})^{\frac{3}{2}} P_0^\mu\,.
\eeq 
It is then easy to show that the general expressions are
\beq
\begin{split}
&P_{2n}^\mu(x_1,\ldots, x_{2n}) = i^{-n^2}(\sigma_{2n}^{(2n)})^{\frac{2n-1}{2}} P_0^\mu\, \\
&P_{2n+1}^\sigma(x_1,\ldots, x_{2n+1}) = i^{n^2+n} (\sigma_{2n+1}^{(2n+1)})^{n} P_1^\sigma\,.
\label{41mu}
\end{split}
\eeq 
Then, by requiring consistency with unperturbed solutions we can fix the constants to $P_0^\mu=1$ and  $P_1^\sigma=1$. 

The $\bal$-dependent part is solved by
\beq
D_{2n}^\mu(x_1,\ldots,x_{2n};\bal)=2^{n}\sqrt{\prod_{i=1}^{2n} \cos\left(\sum_{s \in \mathcal{S}} \frac{\alpha_s}{2}\sum_{j=1}^{2n} \sinh (s \theta_{ij})\right)} \,,
\eeq 
and 
\beq
D_{2n+1}^\sigma(x_1,\ldots,x_{2n+1};\bal)=2^{n} \sqrt{\prod_{i=1}^{2n+1} \cos\left(\sum_{s \in \mathcal{S}}\frac{\alpha_s}{2}\sum_{j=1}^{2n+1} \sinh (s \theta_{ij})\right)} \,.
\eeq 
Putting everything together, 
the form factors of the disorder and order fields are
\beqa
F^\mu_{2n}(\theta_1,\ldots,\theta_{2n}; \bal)=  i^n  \bra \mu \ket_{\bal} \, \sqrt{\prod_{i=1}^{2n} \cos\left(\sum_{s \in \mathcal{S}} \frac{\alpha_s}{2}\sum_{j=1}^{2n} \sinh (s \theta_{ij})\right)} \prod_{i<j} \tanh\frac{\theta_{ij}}{2} \varphi(\theta_{ij};\bal)\,.
\label{solmu}
\eeqa
and 
\beqa
F^\sigma_{2n+1}(\theta_1,\ldots,\theta_{2n+1}; \bal)= i^{n}   
F_1^\sigma(\bal) \sqrt{\prod_{i=1}^{2n+1} \cos\left(\sum_{s \in \mathcal{S}} \frac{\alpha_s}{2}\sum_{j=1}^{2n+1} \sinh (s \theta_{ij})\right)}\prod_{i<j } \tanh\frac{\theta_{ij}}{2} \varphi(\theta_{ij};\bal)\,,
\eeqa
where we used the identities
\beq
\prod_{1\leq i<j\leq 2n} (x_i+ x_j)=\prod_{1\leq i<j\leq 2n} \sqrt{x_i x_j} \, 2 \cosh\frac{\theta_{ij}}{2}
= 2^{n(2n-1)}(\sigma_{2n}^{(2n)})^{\frac{2n-1}{2}} \prod_{1\leq i<j\leq 2n} \cosh\frac{\theta_{ij}}{2}\,. \label{uspro}
\eeq
As we can see from the above expressions, the unperturbed solutions given in \eqref{mu} and \eqref{sigma} are correctly recovered in the $\bal\rightarrow0$ limit. As shown more generally in \cite{PRL}, the form factors of the perturbed theory factorise as
\beqa
F^\mathcal{O}_{n}(\theta_1,\ldots,\theta_n; \bal)= F_{n}^{\mathcal{O}}(\theta_1,\ldots,\theta_{n};\boldsymbol{0}) G_{n}(\theta_1,\ldots,\theta_{n};\bal)\,.
\eeqa
Note that this factorisation property is model-independent. The sole exception is the field $\Theta$ in the Ising field theory.
In the case of $\sigma$ and $\mu$, the deforming factor reads
\beq 
G_{n}(\theta_1,\ldots,\theta_{n};\alpha):= \sqrt{\prod_{i=1}^{n} \cos\left(\sum_{s \in \mathcal{S}} \frac{\alpha_s}{2}\sum_{j=1}^{n} \sinh (s \theta_{ij})\right)} \prod_{i<j}\varphi(\theta_{ij};\bal).
\label{Gn}
\eeq

\subsubsection{Solving the Equations: The Trace of the Stress-Energy Tensor $\Theta$}

The equation for the form factors of the trace of the stress-energy tensor suggests a similar solution procedure as for $\mu$ and $\sigma$. However, a special feature of this field is that it is local, hence its two-particle form factor has no kinematic pole, even if higher-particle ones do. Since their kinematic pole residue is non-vanishing, according to (\ref{thefieldE}), higher particle form factors are non-vanishing in the $\TTb$-perturbed theory. Hence, we need to solve (\ref{thefieldE}) for $n>1$. 
Comparing (\ref{thefieldE}) with (\ref{thefieldmu}) and reviewing the solution to (\ref{thefieldmu}) given by (\ref{solmu}) it is clear that the simplest, consistent solution to the  higher-particle form factor equations will be of the form
\beq 
F_{2n}^{\Theta}(\theta_1,\ldots,\theta_{2n};\bal)=A_{2n}(\bal)\sqrt{\prod_{i=1}^{2n} \sin\left(\sum_{s \in \mathcal{S}} \frac{\alpha_s}{2}\sum_{j=1}^{2n} \sinh (s \theta_{ij})\right)} \prod_{i<j} \tanh\frac{\theta_{ij}}{2}\varphi(\theta_{ij};\bal)\,,\quad n>1
\label{fun}
\eeq 
where $A_{2n}(\bal)$ is a normalisation constant to be fixed later. Taking this solution and writing the  kinematic residue equation that relates the four-particle to the two-particle form factor we get that the latter is given by
\beq 
F_2^\Theta(\theta;\bal)=-A_4(\bal) \left|\sin \left(\sum_{s \in \mathcal{S}} \frac{\alpha_s}{2} \sinh (s \theta)\right)\right| \tanh\frac{\theta}{2}\varphi(\theta;\bal)\,.
\label{66}
\eeq 
This function has no pole at $i\pi$, as required but, unless $A_4(\bal)$ is chosen carefully, it will give zero when the parameters $\alpha_s \rightarrow 0$. One naive way to deal with this is to choose
\beq 
-A_4(\bal)=\frac{2 C(\bal)}{\alpha_{s'}}\,,
\eeq 
and then take the limit $\alpha_{s}$ to zero for each term on the sum over $s$, with $s=s'$ the spin associated with the last limit taken. This limit then gives
\beq 
\lim_{\alpha_{s'}\rightarrow 0} \lim_{\alpha_{s\neq s'}\rightarrow 0} F_2^\Theta(\theta;\bal)=F_2^\Theta(\theta;\bol)=C(0)\left|\sinh\frac{s'\theta}{2}\right|\tanh \frac{\theta}{2}\,.
\eeq 
This result is problematic. Not only does is not reproduce the correct form factor at $\bal=0$ but is dependent on the order in which the parameters $\alpha_s$ are taken to zero, a property that seems quite unnatural. We must then conclude that the ansatz (\ref{fun}) is not the correct solution to the form factor equations for $n>1$. The question is then, how should (\ref{fun}) be modified so that it is consistent with a two-particle form factor which has all the correct properties?

Given that the oscillatory part of the solution (\ref{fun}) is crucial for the solution of the higher particle form factors, it seems that we must accept the presence of an oscillatory part also in the two-particle form factor. We assume then the following form
\beq 
F_2^\Theta(\theta;\bal)=2\pi f(\theta;\bal) \frac{F_{\rm min}(\theta;\bal)}{F_{\rm min}(i\pi;\bal)}\,,
\eeq 
where $f(\theta;\bal)$ is a function that satisfies
\beq 
f(i\pi;\bal)=f(\theta;\bol)=1\,,
\eeq 
as well as being symmetric and $2\pi i$-periodic in $\theta$. According to the discussion above, it must also include the oscillatory factor in (\ref{fun}), specialised to $n=1$. It is not too difficult to realise that the function we need is
\beq 
f(\theta;\bal)=\left|\frac{\sin \left(\sum_{s \in \mathcal{S}} \frac{\alpha_s}{2} \sinh (s \theta)\right)}{\sum_{s \in \mathcal{S}} \frac{\alpha_s}{2} \sinh (s \theta)}\right|\cosh\frac{\theta}{2}\,,
\eeq 
which gives the two-particle form factor
\beq 
F_2^\Theta(\theta;\bal)=-2\pi i \left|\frac{\sin \left(\sum_{s \in \mathcal{S}} \frac{\alpha_s}{2} \sinh (s \theta)\right)}{\sum_{s \in \mathcal{S}} \frac{\alpha_s}{2} \sinh (s \theta)}\right| \sinh\frac{\theta}{2} \varphi(\theta;\bal).
\label{sol}
\eeq 
Comparing (\ref{sol}) to (\ref{66}) we see that we need to modify the ansatz (\ref{fun}) so as to produce the function (\ref{sol}). In order to do this, while still solving the higher-particle form factor equations we need to find a function $f_{2n}(x_1,\ldots, x_{2n};\bal)$
such that we can multiply (\ref{fun}) by it, while still solving all equations. In other words, we need this function to satisfy
\beq 
f_{2n+2}(-x,x,x_1,\ldots,x_{2n};\bal)=f_{2n}(x_1,\ldots,x_{2n};\bal)\,.
\label{propf}
\eeq
and 
\beq 
f_2(x_1,x_2;\bal)=\frac{\cosh\frac{\theta_{12}}{2}}{ \left| \sum_{s \in \mathcal{S}}\frac{\alpha_s}{2}\sinh(s\theta_{12}) \right|}\,.
\label{rat}
\eeq 
A crucial observation to proceed is that the ratio of elementary symmetric polynomials
\beq 
\frac{\sigma_1^{(2n)} \sigma^{(2n)}_{2n-1}}{\sigma^{(2n)}_{2n}}\,,
\eeq
solves (\ref{propf}). This is because under the reduction $(x_1,x_2,\ldots,x_{2n+2})\mapsto (-x,x,x_1,\ldots,x_{2n})$, the elementary symmetric polynomials behave as follows: $\sigma^{(2n+2)}_{2n+1}\mapsto -x^2 \sigma^{(2n)}_{2n-1}$,  $\sigma^{(2n+2)}_{2n+2}\mapsto -x^2\sigma^{(2n)}_{2n}$ and $\sigma^{(2n+2)}_1 \mapsto \sigma^{(2n)}_1$\,. In fact, any function of the ratio (\ref{rat}) will solve (\ref{propf}). Additionally, this facts will still hold true if the elementary symmetric polynomials are taken to be functions of the powers $x_i^s$ rather than just $x_i=e^{\theta_i}$. Let us then be slightly more general and denote as $\sigma_{i,s}^{(2n)}$
the ith elementary symmetric polynomial of the $2n$ variables $\{e^{s\theta_1}, \ldots, e^{s\theta_{2n}}\}$. It is a matter of simple algebra to verify that for any odd integer values of $s$, the function
\beq 
g_{2n}(x_1,\ldots,x_{2n};s)=\frac{\sigma^{(2n)}_{1,s} \sigma^{(2n)}_{2n-1,s}}{\sigma^{(2n)}_{2n,s}}\,,
\eeq 
solves (\ref{propf}).
We now proceed to write (\ref{rat}) in terms of these factors. For $n=1$ we have that
\beq 
g_2(x_1,x_2;s)=\frac{(\sigma^{(2)}_{1,s})^2}{\sigma^{(2)}_{2,s}}=\frac{(e^{s\theta_1}+e^{s\theta_2})^2}{e^{s\theta_1+s\theta_2}}=4\cosh^2\frac{s\theta_{12}}{2}\,.
\eeq 
so that
\beq 
4\sinh^2\frac{s\theta_{12}}{2}=g_2(x_1,x_2;s)-4\,.
\eeq 
and 
\beq 
\sqrt{g_2(x_1,x_2;s) (g_2(x_1,x_2;s)-4)}=2|\sinh(s\theta_{12})|\,.
\label{abssinh}
\eeq 
This is precisely the sort of factor we need in our two-particle form factor. In fact, in order to have a consistent set of form factors, we need to multiply our solution (\ref{fun}) by the function
\beq 
f_{2n}(x_1,\ldots,x_{2n};\bal)=\frac{\sqrt{\frac{\sigma^{(2n)}_{1}\sigma^{(2n)}_{2n-1}}{\sigma^{(2n)}_{2n}}}}{\left|\sum\limits_{s\in\mathcal{S}}\frac{\alpha_s}{4}\sqrt{\left(\frac{\sigma^{(2n)}_{1,s} \sigma^{(2n)}_{2n-1,s}}{\sigma^{(2n)}_{2n,s}}-4\right)\frac{\sigma^{(2n)}_{1,s} \sigma^{(2n)}_{2n-1,s}}{\sigma^{(2n)}_{2n,s}}}\right|}\,,\label{666}
\eeq 
where $\sigma_i^{(2n)}\equiv \sigma_{i,1}^{(2n)}$ and we set $f_0(\bal)=1$.
For $n=1$ this function correctly reproduces (\ref{rat}) and the two-particle form factor reduces to (\ref{sol}) with an appropriate choice of normalisation. This requires that $|\sum_{s\in\mathcal{S}} \frac{\alpha_s}{2}\sinh(s\theta)|=|\sum_{s\in\mathcal{S}} \frac{\alpha_s}{2}|\sinh(s\theta)||$ which holds for $\theta\in\mathbb{R}$.

Let us then summarise what we found. All the form factors of the field $\Theta$, with the standard normalisation $F_0=F_2^\Theta(i\pi)=2\pi$, are given by 
\beqa 
F_{2n}^{\Theta}(\theta_1,\ldots,\theta_{2n};\bal)&=& 2\pi i \,i^{n}
\frac{\sqrt{\prod\limits_{i=1}^{2n} \sin\left(\sum\limits_{s \in \mathcal{S}} \frac{\alpha_s}{2}\sum\limits_{j=1}^{2n} \sinh (s \theta_{ij})\right)}}{\left|\sum\limits_{s\in\mathcal{S}}\frac{\alpha_s}{4}\sqrt{\left(\frac{\sigma^{(2n)}_{1,s} \sigma^{(2n)}_{2n-1,s}}{\sigma^{(2n)}_{2n,s}}-4\right)\frac{\sigma_{2n}^{(2n)}\sigma^{(2n)}_{1,s} \sigma^{(2n)}_{2n-1,s}}{\sigma^{(2n)}_{2n,s}\sigma^{(2n)}_{1} \sigma^{(2n)}_{2n-1}}}\right|}\nonumber\\
&& \times \prod_{i<j} \tanh\frac{\theta_{ij}}{2}\varphi(\theta_{ij};\bal)\,.
\label{fun2}
\eeqa
Note that, although the final formula contains $\bal$-dependence both in the numerator and denominator, it is easy to see $F_{2n}^{\Theta}(\theta_1,\ldots,\theta_{2n};\bol)=0$ for $n>1$.

\section{Two-Point Function Cumulant Expansion}
Let us consider a generic field $\mathcal{O}$. We are going to employ the form factors found in the previous section to write the cumulant expansion of the two-point function. For fields with non-vanishing VEV we can write
\beq 
\log\frac{\bra \mathcal{O}(0)\mathcal{O}(r)\ket_{\bal}}{\bra \mathcal{O} \ket^2_{\bal}}=\sum_{\ell=1}^\infty c_{\ell}^\mathcal{O} (r;\bal)\,,
\label{cumulant}
\eeq 
where $c_{\ell}^\mathcal{O} (r;\bal)$ are the cumulants,
\beqa 
c_{\ell}^\mathcal{O} (r;\bal)&:=&\frac{1}{\ell!}\int_{-\infty}^\infty \frac{d\theta_1}{2\pi} \cdots \int_{-\infty}^\infty \frac{d\theta_{\ell}}{2\pi} e^{-r\sum_{i=1}^{\ell} \cosh\theta_i} h_{\ell}^\mathcal{O}(\theta_1,\ldots, \theta_{\ell};\bal)\nonumber\\
&=& \frac{1}{\pi \ell!}\int_{-\infty}^\infty \frac{d\theta_1}{2\pi} \cdots \int_{-\infty}^\infty \frac{d\theta_{\ell-1}}{2\pi} K_0(r d_\ell) h_{\ell}^\mathcal{O}(\theta_1,\ldots, \theta_{\ell-1},0;\bal)\,.
\label{cumu2}
\eeqa 
Here we introduced the notation 
\beq 
d_\ell=\sqrt{\left(\sum_{j=1}^{\ell-1} \cosh\theta_i+1\right)^2-\left(\sum_{j=1}^{\ell-1} \sinh\theta_i\right)^2}\,,
\label{dell}
\eeq 
and $K_0(x)$ is the modified Bessel function. The second line of \eqref{cumu2} follows from integration of one rapidity variable, since the functions $h_\ell(\{\theta_i\};\bal)$ depend on rapidity differences only. This expression for the cumulants can be found in many places, for instance \cite{BK}.

In the Ising model, due to $\mathbb{Z}_2$ symmetry, we usually have either even- or odd-particle cumulants only. For instance for the field $\mu$
\beqa 
\bra \mu\ket^2_{\bal} h_2^\mu(\theta_1,\theta_2;\bal)&:=&|F_2^\mu(\theta_1,\theta_2;\bal)|^2\,,\\
\bra \mu\ket^4_{\bal} h_4^\mu(\theta_1,\theta_2,\theta_3,\theta_4;\bal)&:=&\bra \mu\ket_{\bal}^2 |F_4^\mu(\theta_1,\theta_2,\theta_3,\theta_4;\bal)|^2-|F_2^\mu(\theta_1,\theta_2;\bal)|^2|F_2^\mu(\theta_3,\theta_4;\bal)|^2\nonumber\\
&& - |F_2^\mu(\theta_1,\theta_3;\bal)|^2|F_2^\mu(\theta_2,\theta_4;\bal)|^2\nonumber\\
&& - |F_2^\mu(\theta_1,\theta_4;\bal)|^2|F_2^\mu(\theta_2,\theta_3;\bal)|^2\,,
\eeqa 
whereas for $\Theta$ we just need to replace $\mu \mapsto \Theta$ and $\bra \mu \ket_{\bal} \mapsto 2\pi $. For the field $\sigma$ on the other hand, only the odd cumulants are not vanishing, the first two of which are
\beqa 
h_1^\sigma(\theta_1;\bal)&:=& 1\,,\\
|F_1^\sigma(\bal)|^6 h_3^\sigma(\theta_1,\theta_2,\theta_3;\bal)&:=&|F_1^\sigma(\bal)|^4|F_3^\sigma(\theta_1,\theta_2,\theta_3;\bal)|^2-|F_1^\sigma(\bal)|^6\,.
\eeqa 
Note that, since the $1$-point function of $\sigma$ vanishes, the normalisation in the cumulant expansion \eqref{cumulant} should be replaced by the norm squared of the one-particle form factor
\beq
    \log\frac{\bra\sigma(0)\sigma(r)\ket_{\bal}}{\vert F_1^{\sigma}(\bal)\vert^2} = \sum_{\ell = 1}^{\infty} c_{\ell}^{\sigma}(r;\bal)\;.
\eeq
Under a generalised $\mathrm{T}\overline{\mathrm{T}}$ perturbation, a free theory will cease to be such, in the sense that many of the nice properties free theories enjoy -- such as a trivial $S$-matrix -- are lost to the deformation. In particular, the cumulant expansion does not simplify in any obvious way, a fact which is well-known to happen for the Ising fields in the usual free model \cite{Yurov:1990kv} and on its replica version, studied in the context of entanglement measures (see i.e. \cite{nexttonext, lattest}). These simplifications are generally due to the Pfaffian structure of the form factors that follows from Wick's theorem in free theories. Such properties no longer hold in generalised $\mathrm{T}\overline{\mathrm{T}}$ perturbations.  Instead, as far as we can see, with our form factor solutions, there is no simple closed general expression for the cumulants. It is however interesting to write down at least the lower ones. As we shall see, they exhibit interesting properties, some of which will extend to the whole cumulant expansion. 

\subsection{Two-Particle Cumulants and their Asymptotics}
Let us look at the lower-particle contributions to the expansion for the fields $\Theta$ and $\mu$. For the field $\mu$ we have
\beqa 
c_2^\mu(r;\bal)&=&\frac{1}{2}\int_{-\infty}^\infty \frac{d\theta_1}{2\pi} \int_{-\infty}^\infty \frac{d\theta_2}{2\pi} e^{-2r \cosh\frac{\theta_{12}}{2} \cosh \frac{\hat{\theta}_{12}}{2}} \cos^2\left(\sum_{s\in \mathcal{S}}\frac{\alpha_s}{2}\sinh(s\theta_{12})\right)\nonumber\\
&& \times \prod_{s\in \mathcal{S}}e^{\frac{\alpha_s \theta_{12}}{\pi}\sinh(s\theta_{12})}\tanh^2\frac{\theta_{12}}{2}\,.
\eeqa 
As usual, one integral can be carried out by introducing new variables $x=\theta_{12}$ and $y=\hat{\theta}_{12}:=\theta_1+\theta_2$, so the cumulant becomes
\beq 
c_2^\mu(r;\bal)=\frac{1}{4\pi^2}\int_{-\infty}^\infty  dx  K_0(2r \cosh\frac{x}{2}) \cos^2\left(\sum_{s\in \mathcal{S}}\frac{\alpha_s}{2}\sinh(s x)\right) \prod_{s\in \mathcal{S}}e^{\frac{\alpha_s x}{\pi}\sinh(s x)}\tanh^2\frac{x}{2}\,.
\label{cumu}
\eeq 
A very similar behaviour is found in the first cumulant contribution to the two-point function of $\Theta$. Following the same steps, we arrive at
\beq 
c_2^\Theta(r;\bal)=\frac{1}{4\pi^2}\int_{-\infty}^\infty  dx \, K_0(2r \cosh\frac{x}{2}) \left[\frac{\sin\left(\sum_{s\in \mathcal{S}}\frac{\alpha_s}{2}\sinh(s x)\right)}{\sum_{s\in \mathcal{S}}\frac{\alpha_s}{2}\sinh(s x)} \right]^2 \prod_{s\in \mathcal{S}}e^{\frac{\alpha_s x}{\pi}\sinh(s x)} \sinh^2\frac{x}{2}\,.
\label{cuthe}
\eeq 
Regarding the convergence of the integrals \eqref{cumu} and \eqref{cuthe} it is easy to see that this is determined by the exponential factor and, in particular, by the sign of the parameter $\alpha_{s^*}$ where $s^*$ is the largest spin which is present in the product of exponentials. If $\alpha_{s^*}>0$ the integral is divergent, whereas the opposite is true for $\alpha_{s^*}<0$. This applies not just to the second cumulants, but to all cumulants. The same conclusion is reached by employing a well-known derivation presented in \cite{DelB1,DelB2}, which concerns the asymptotic behaviour of form factors and correlation functions. In these papers it was shown that for a two-point function $\bra \mathcal{O}(0) \mathcal{O}^\dagger(r) \ket$ to admit a convergent form factor expansion which is compatible with power-law scaling at short distances, the form factors of $\mathcal{O}$ can diverge at the very most exponentially with each rapidity variable, that is
\beq 
F^{\mathcal{O}}_k(\theta_1,\ldots,\theta_k;\bol) \sim  e^{y_{\mathcal{O}} \theta_i}\quad \mathrm{for}\quad \theta_i \rightarrow \infty,
\label{bound}
\eeq  
with $y_{\mathcal{O}} \leq \Delta_{\mathcal{O}}$ and where $\Delta_{\mathcal{O}}$ is the conformal dimension of the field in the UV limit. 

As we have seen, in generalised $\TTb$-perturbed theories the asymptotics of the form factors is dictated by the double-exponential function $\exp({\frac{\alpha_{s^*}}{\pi}\theta\sinh(s\theta)})$. The remaining $\bal$-dependent part of the form factor is generally a product of real oscillatory functions, which are bounded\footnote{This is slightly different for the field $\Theta$ as we have seen, but does not substantially change our argument.} and the ``unperturbed" form factor will typically scale exponentially.  Clearly, when $\alpha_{s^*}>0$ the form factors diverge double-exponentially so the bound (\ref{bound}) can never be satisfied and the expansion of the two-point function is divergent. However, when $\alpha_{s^*}<0$ the bound (\ref{bound}) is always satisfied, since the form factors tend to zero for large rapidities. Thus we expect power-law scaling which we indeed also find from the cumulant expansion. 

Notice that these convergence properties chime with what was mentioned in the introduction: the TBA for generalised $\TTb$ deformations with more than one coupling are only well defined if $\alpha_{s^*}<0$.  We must mention however, that the asymptotics of these expressions and higher terms in the form factor expansion can change substantially if the set of spins $s$ involved in the products above is infinite, as it will now depend on the analytic properties of the infinite product of exponentials. For the purposes of this paper and in order to avoid such subtleties, we will always assume that we are considering a finite set of non-vanishing couplings, even if the total set $\mathcal{S}$ is infinite.

\subsection{Leading Cumulant for the $\mathrm{T}\overline{\mathrm{T}}$ Perturbation}

 Let us now focus on the standard $\mathrm{T}\overline{\mathrm{T}}$ perturbation corresponding to having a single non-vanishing parameter $\alpha_1:=\alpha$. Other more general cases are discussed in Appendix A. 

\subsubsection{$\alpha>0$: Excitations of Finite Positive Length}
According to the picture put forward in \cite{Cardy:2020olv} the $\alpha>0$ regime corresponds to a theory whose UV physics is characterised by objects of finite length. It is then not surprising that short distances cannot be probed and that in this limit the two-point function should diverge. At the TBA level \cite{Cavaglia:2016oda} this finite length leads to a finite ground state energy.

In fact, for $\alpha>0$, even the IR limit can be problematic, if we also probe high momenta. In this case we will again encounter this finite length, leading to a divergent form factor expansion. In other words, there is no local QFT describing the deep UV limit of the theory and if we try to probe this theory by approaching either small values of $r$ for any momenta, or large values of $r$ and high momenta we always find a divergence. For large momenta, we can even estimate a characteristic scale or cut-off -- a function of $r/\alpha$ -- at which this divergence dominates as we now show.

If we take the two-particle cumulant of $\mu$ and  expand the Bessel function at leading order for large $r$ we find
\beq 
{c}_2^\mu(r\gg 1;\alpha>0)\approx \frac{1}{4\pi^2}\sqrt{\frac{\pi}{2 r}}\int_{-\infty}^\infty  \frac{dx}{\sqrt{\cosh\frac{x}{2}}}  \cos^2\left(\frac{\alpha}{2}\sinh x\right) e^{-2r \cosh\frac{x}{2}+\frac{\alpha x}{\pi}\sinh x}\tanh^2\frac{x}{2}\,.
\eeq 
Convergence of the integral is determined by the balance between the two terms in the exponent. While for $\alpha<0$ the exponent is negative and the exponential rapidly decreasing, for $\alpha>0$ the integral is convergent for as long as
\beq 
2r \cosh\frac{x}{2} > \frac{\alpha x}{\pi} \sinh x\,.
\label{comp}
\eeq
In other words, there exist a maximum allowed rapidity $x$, solution of the equation
\beq  
\frac{\pi r}{\alpha}=x \sinh\frac{x}{2}\,.
\eeq 
For $x\gg 0$ we can approximate the $\sinh$ by an exponential, arriving at 
\beq 
\frac{\pi r}{\alpha}= \frac{x}{2}  e^{\frac{x}{2}}\,.
\label{71}
\eeq 
The solutions to the equation above are known in the literature as Lambert $W$-function $W_k(z)$ for $k\in \mathbb{Z}$. This is a multi-valued function, hence the index $k$. For real values of $x$ and $\frac{\pi r}{\alpha}>0$ the correct branch of the solution is the principal one $x=2 W_0(\pi r/\alpha)$. This scales as
\beq  
W_0(a) \approx a \quad {\rm for} \quad  a\ll 1\,,\quad {\rm and} \quad
W_0(a) \approx \log a \quad {\rm for} \quad  a\gg 1\,,
\eeq 
Quite naturally, the cut-off depends on the ratio of the two independent scales in the system $r/\alpha$. If $r$ is large compared to $\alpha$ this cut-off will be very large as well since the finite length of the fundamental excitation is not seen, even for large momenta. However, if $\alpha$ is large compared to $r$ the particles' finite length is probed for momenta that are not necessarily large and the cumulant expansion can only be made convergent by the introduction of the rapidity cut-off 
\beq 
\Lambda=2W_0(\pi r/\alpha)\,.
\label{cutoff}
\eeq 
The fact that Lambert's $W$-function plays a role is rather suggestive given that it features in a variety of problems, many of which are intuitively connected to the physics of $\TTb$ perturbations (see \cite{waterloo} for a review). For instance, a key application is to delay problems which are similar in spirit to the ``delay" induced by the phase-shift $\Phi_{\bal}(\theta)$ and by the finite length of the rods, that best describe the effect of the $\TTb$ perturbation in the GHD context. The Lambert function also features directly in previous discussions of $\TTb$ perturbations, such as in \cite{Cardy:2020olv} where it enters the formula for the free energy in a hard-rod model and in the solutions of $\mathrm{JT}$ gravity \cite{Dubovsky:2017cnj,MannOh}.

Although the characteristic momentum cut-off \eqref{cutoff} is obtained from the two-particle contribution, the qualitative picture presented here should extend to higher-particle terms. It is indeed possible to write a simple argument showing this. Consider the definition \eqref{cumu2} in the large $r$ limit. Expanding the Bessel function, for large $r$, all cumulants will include an $r$-dependent exponential of the form $e^{-r d_\ell}$
with $d_\ell$ defined in \eqref{dell}. Suppose that one of the rapidities is very large, say $\theta_1\gg 1$ and $\theta_1 \gg \theta_{i\neq 1}$. Then, $\cosh\theta_1, \sinh\theta_1$ are also much larger than $\cosh \theta_{i\neq 1}$ and $\sinh \theta_{i\neq 1}$. By neglecting these terms in the definition of $d_{\ell}$, we have the approximation
\beq
e^{-r d_\ell} \approx e^{-2 r \cosh\frac{\theta_1}{2}} \quad \mathrm{for} \quad \theta_1\gg 1, \, \theta_1\gg \theta_{i\neq 1}\,. 
\eeq 
The cumulant will also contain the factor
\beq 
\prod_{1\leq i \leq j \leq \ell-1} e^{\frac{\alpha \theta_{ij}}{\pi} \sinh \theta_{ij}}\prod_{i=1}^{\ell-1} e^{\frac{\alpha \theta_{i}}{\pi} \sinh \theta_{i}}\approx e^{-\frac{\ell \alpha \theta_1}{\pi} \sinh\theta_1}\,,
\eeq 
where the approximation is made under the same assumptions as above. Then the two exponentials can be compared just as for the two-particle cumulant and a similar cut-off $\Lambda$ -- with $\alpha$ replaced by $\ell\alpha$ -- can be found.
\medskip
\subsubsection{$\alpha<0$: Excitations of Finite ``Negative" Length}
Following the analysis of \cite{Cardy:2020olv}, for $\alpha<0$ we are now dealing with a theory of particles having a ``negative" length. From the viewpoint of correlation functions, both the UV and IR regimes can be probed without any issues at all. 
In fact, the form factor series is more rapidly convergent for $\alpha<0$ than for $\alpha=0$ since the minimal form factor typically decays much faster for large rapidities than any other function in the integrand. We have that, for all cumulants in \eqref{cumu2}
\beq
\lim_{\theta_i\rightarrow \infty} h_\ell^{\mathcal{O}}(\theta_1,\ldots,\theta_\ell;\alpha)=0\quad \forall \, i\, \& \, \alpha<0\,.
\label{eq:h_to_0}
\eeq 
For many theories, the statement above also holds at $\alpha=0$ but the Ising model is an exception to this rule since the cumulants of $\mu, \sigma$ and $\Theta$ all either diverge exponentially  ($\Theta$) or tend to a constant ($\mu$ and $\sigma$) for large rapidities. 

Now that \eqref{eq:h_to_0} holds, we can safely expand the Bessel functions $K_0(2r d_\ell)$ in \eqref{cumu2} for small $r$ in order to investigate the UV scaling of correlators,
\beq 
K_0(r d_\ell)=-\log r -\gamma + \log 2 -\log d_\ell\,.
\eeq 
We find that the leading contributions to the cumulant expansion at short distances are all proportional to a factor $\log r$. This implies a power-law scaling of the correlators, just as one would expect if the UV limit was a CFT. This is an interesting though unexpected result, given that we know a $\TTb$-deformed theory is not supposed to have a UV fixed point in the standard sense. Our results suggest that for large energies, the $\TTb$ deformation does have properties that are reminiscent of critical systems. In other words, two-point functions will still scale as power-laws $r^{-4z^{\mathcal{O}}(\alpha)}$ at short distances, with exponent $z^{\mathcal{O}}(\alpha)<\Delta^{\mathcal{O}}$ -- for $\alpha<0$ -- where $\Delta^{\mathcal{O}}$ is the UV dimension of the field in the unperturbed theory. We also see that $z^{\mathcal{O}}(\alpha) \rightarrow 0$ for $-\alpha \rightarrow \infty$. The IR limit is straightforwardly analysed as well, with two-point functions decaying exponentially at large distances, faster than in the undeformed theory. In the deep IR the theory will still be trivial, as dictated by the presence of a mass scale. 
\medskip

For the two-particle cumulants, the leading UV ($r\ll 1$) contributions are
\beq 
c_2^\mu(r\ll 1;\alpha<0)\approx-\frac{\log r}{4\pi^2}\int_{-\infty}^\infty  dx   \cos^2\left(\frac{\alpha}{2}\sinh x\right) e^{\frac{\alpha x}{\pi}\sinh x}\tanh^2\frac{x}{2}:=-4z_2^\mu(\alpha)\log r\,,
\label{c2mu}
\eeq 
and similarly
\beq 
c_2^\varepsilon(r\ll 1;\alpha<0)\approx-\frac{\log r}{4\pi^2}\int_{-\infty}^\infty  dx  \left[\frac{\sin(\frac{\alpha}{2}\sinh x)} {{\alpha}\cosh \frac{x}{2}}\right]^2 e^{\frac{\alpha x}{\pi}\sinh x}:=-4z_2^\varepsilon(\alpha)\log r\,,
\label{c2ep}
\eeq 
The cumulants of the energy field $\varepsilon$ which is related to the $\Theta$ by $\Theta=2\pi m \varepsilon$ are identical to those of $\Theta$ when $m=1$, however, unlike $\Theta$, $\varepsilon$ is a primary field in CFT and should scale as $1/r^2$ in the short-distance limit for $\alpha=0$ (whereas $\Theta$ vanishes in the CFT limit). It is thus more interesting to consider the short-distance scaling of $\varepsilon$ than that of $\Theta$.

\begin{figure}[h!]
\begin{center}
	\includegraphics[width=7.5cm]{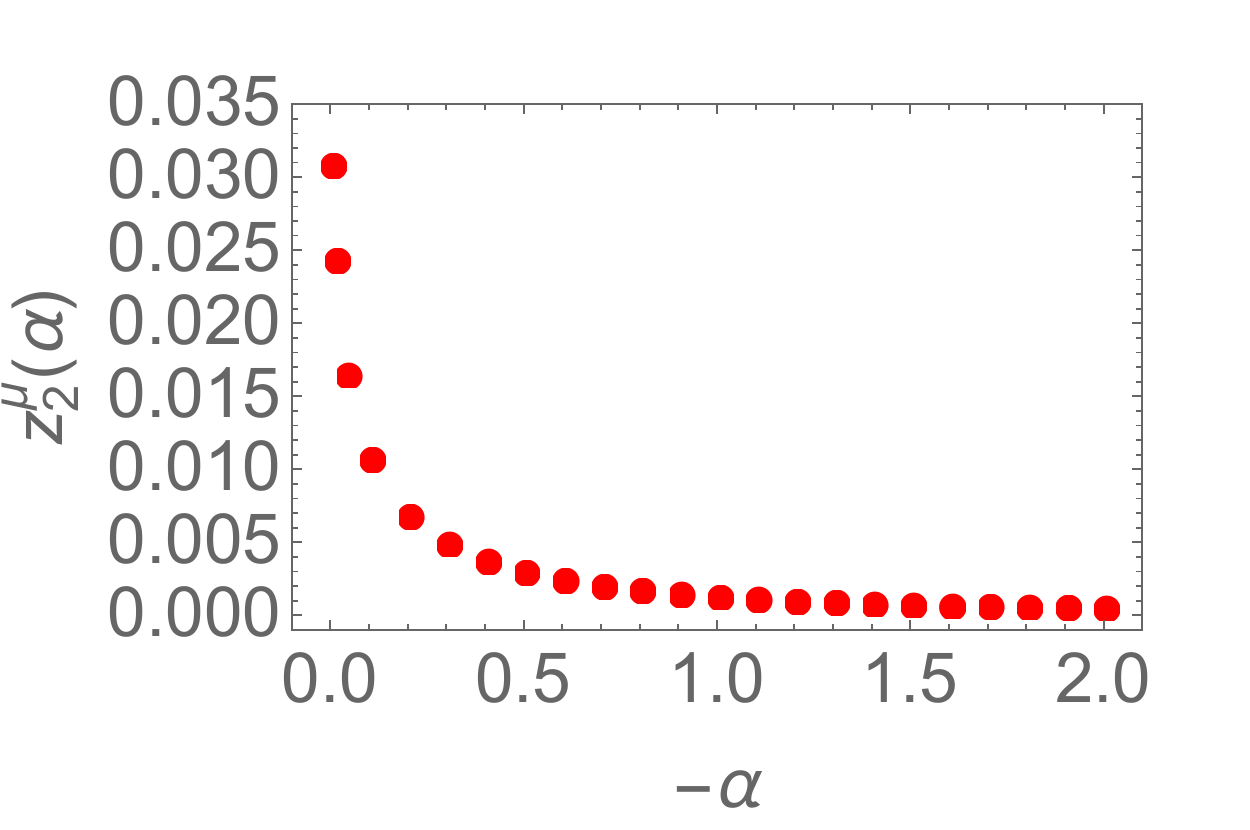}
 \includegraphics[width=7.5cm]{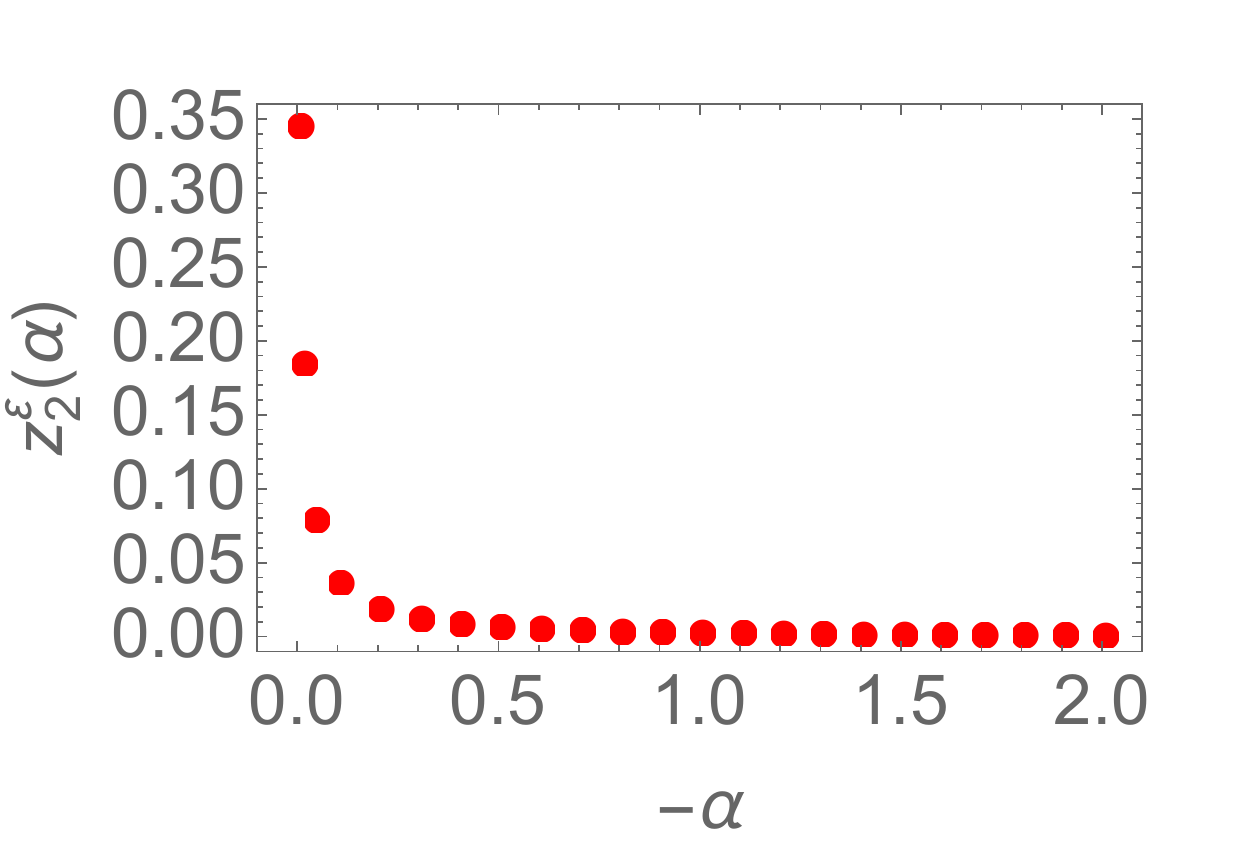}
				    \caption{The functions $z^\mu_2(\alpha)$ and $z_2^\varepsilon(\alpha)$. In the unperturbed Ising model we have that the two-point functions scale exactly as $r^{-\frac{1}{4}}$ and $r^{-2}$ for $\mu$ and $\varepsilon$ respectively so that  $z_2^\mu(0)= \frac{1}{16}$ and $z_2^\varepsilon(0)=\frac{1}{2}$. From the two-particle contributions, it seems that there is still power-law scaling at short distances but the decay becomes faster the larger is $-\alpha$. Note that the definitions \eqref{c2mu} and \eqref{c2ep} of $z^\mu_2(\alpha)$ and $z_2^\varepsilon(\alpha)$ are not valid at $\alpha=0$: there is a sharp change at $\alpha=0$ which is a peculiar feature of the Ising model.}
				     \label{figure1}
    \end{center}
    \end{figure}

Due to the exponential terms, the integrals in \eqref{c2mu} and \eqref{c2ep} cannot be computed analytically but they can be evaluated numerically and the results are shown in Fig.~\ref{figure1}. These indicate that the short-distance scaling of the two-point function will now be dependent on $\alpha$. Note that it is not possible to extrapolate these contributions to $\alpha=0$ because the integral is not convergent in this case (see \cite{Yurov:1990kv}). This sharp change in the behaviour of $z_2^{\mu,\varepsilon}(\alpha)$ is a special feature of the Ising model. 

\subsection{$\Delta$-Sum Rule and $c$-Theorem}
In this section we investigate whether two standard IQFT formulae typically employed a consistency checks for form factor computations can still provide some insights in $\TTb$ perturbed theories.  These are the $\Delta$-sum rule \cite{DSC} and Zamolodchikov's $c$-theorem \cite{cZam}. Once more we limit our consideration to the simplest case of the $\TTb$ deformation. 

Given an IQFT resulting from the massive perturbation of a known CFT, the $\Delta$-sum rule \cite{DSC}  reads
\begin{equation}
   \Delta_{\mathcal{O}}^{UV} - \Delta_{\mathcal{O}}^{IR}= - \frac{1}{4 \pi \langle \mathcal{O} \rangle} \int_0^\infty dr\, r \langle \Theta(0) \mathcal{O}(r) \rangle_c,
\end{equation}
where $\mathcal{O}(x)$ is a local field  in the off-critical theory, $\Delta_{\mathcal{O}}^{IR}=0$ if the theory is massive and the index `c' indicates the connected correlator. 

\begin{figure}[h!]
\begin{center}
	\includegraphics[width=7.5cm]{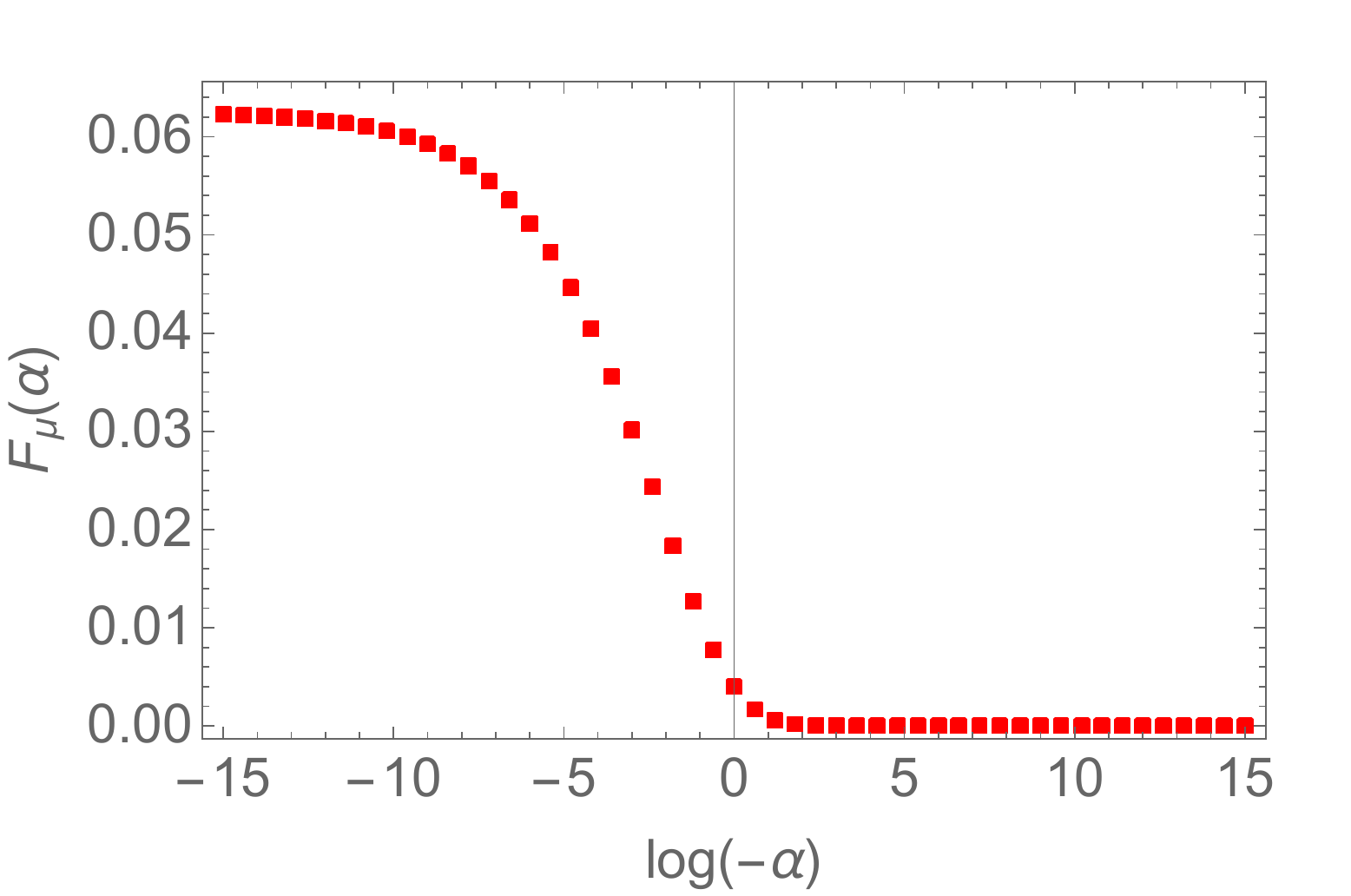}
 \includegraphics[width=7.5cm]{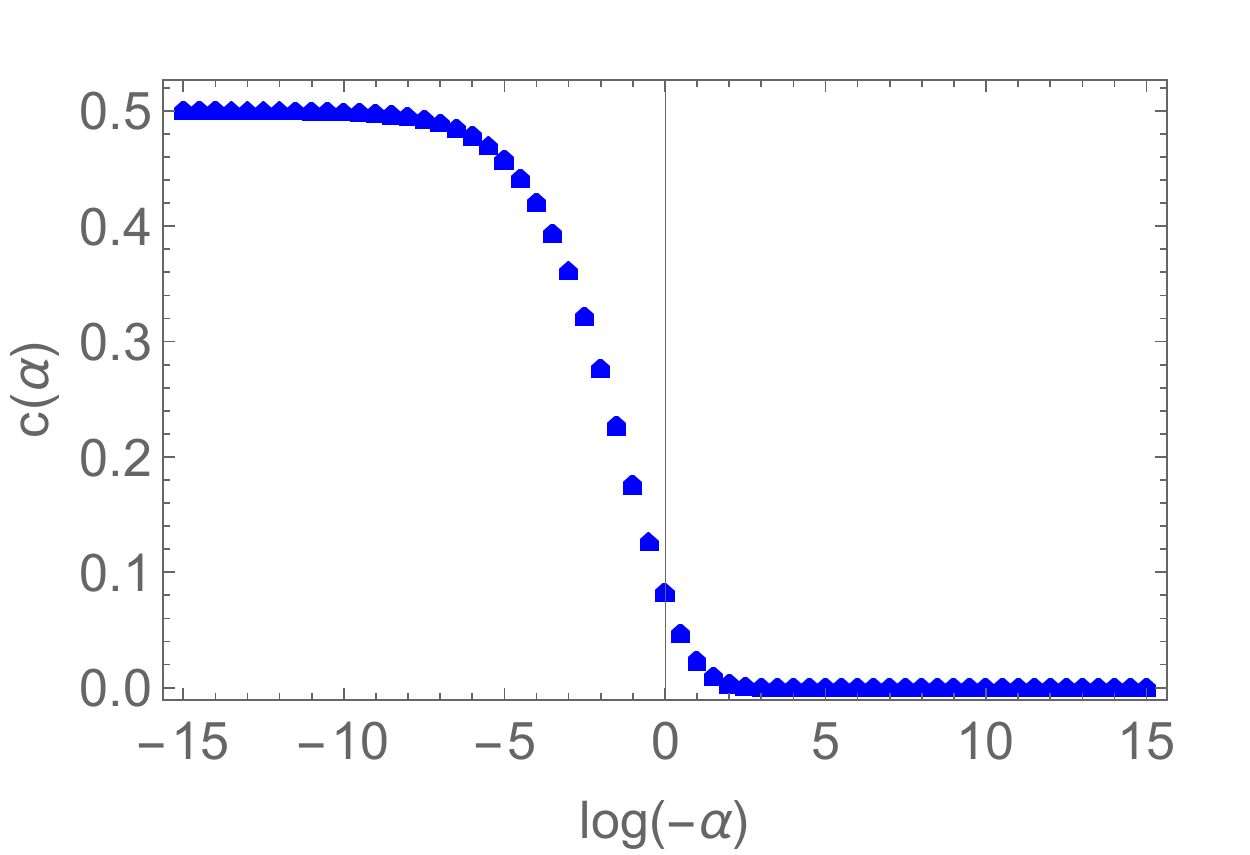}
\caption{The functions $F(\alpha)$ and $c(\alpha)$. They both tend to zero for large $-\alpha$ and to their exact values in the unperturbed theory for $\alpha=0$: $F(0)=\frac{1}{16}=0.0625$ and $c(0)=\frac{1}{2}=0.5$}
				     \label{figure2}
    \end{center}
    \end{figure}

Consider the field $\mu$. After integrating in $r$ and in one of the rapidities, we obtain the exact formula
\begin{equation}
    F_\mu(\alpha):=\Delta_{\mu}^{UV} - \Delta_{\mu}^{IR} = \frac{1}{16 \pi} \int_{ - \infty}^{+ \infty} d x \,\left|\frac{\sin({\alpha}\sinh x)}{\alpha \sinh x} \right| \frac{\tanh^2{\frac{x}{2}}}{\cosh{\frac{x}{2}}}\, e^{\frac{\alpha}{\pi} x \sinh{x}}\,,
\end{equation}
For $\alpha=0$ the integral can be computed exactly and we obtain the expected CFT value $\Delta_{\mu}=\frac{1}{16}$ while for $\alpha < 0$ we can only integrate numerically. We find that $F(\alpha)$ tends to zero for $-\alpha \rightarrow \infty$. The decay is extremely fast and can be best seen when plotting against $\log(-\alpha)$, as in Fig.~\ref{figure2} (left). The behaviours at small and large $-\alpha$ are exactly what we would expect from the general picture developed in the analysis of the cumulants, thus the results of the $\Delta$-sum rule are compatible with the general picture of power-law scaling at short distances.  
    
In the same way we can study  Zamolodchikov's $c$-function. We know that
\begin{equation}
    c^{UV} - c^{IR}=  \frac{3}{2} \int_0^{\infty} dr \, r^3 \langle \Theta(0) \Theta(r) \rangle_c,
\end{equation}
and we can once again integrate in $r$ and in one rapidity to obtain the formula
\begin{equation}
c(\alpha):= c^{UV} - c^{IR} \approx \frac{3}{8} 
     \int_{ - \infty}^{+ \infty} dx \left[\frac{\sin(\frac{\alpha}{2}\sinh x)}{\frac{\alpha}{2} \sinh x} \right]^2\frac{{\tanh^2{\frac{x}{2}}}}{\cosh^2{\frac{x}{2}}} e^{\frac{\alpha}{\pi}x \sinh{x}}\,,
\end{equation}
in the two-particle approximation.
For $\alpha=0$ we obtain the usual value $c=1/2$ while $c(\alpha) \rightarrow 0$ for $\alpha \rightarrow -\infty$. 
The decay to zero as $-\alpha$ is increased is very fast and best seen in logarithmic scale, as shown in Fig.~\ref{figure2} (right).

\section{Conclusions, Discussion and Outlook}
\label{conclusion}
In this paper and, more generally, in the letter \cite{PRL} we have shown that a consistent form factor program can be developed for fields in generalised $\mathrm{T}\overline{\mathrm{T}}$-perturbed integrable quantum field theories. This means that form factor equations can be written in the usual way but they will now be dependent on a scattering matrix which contains additional CDD factors as shown in \eqref{2}. The result are equations that are generally harder to solve and whose solutions have an unusual dependence on the rapidities (typically double-exponential and oscillatory). In particular, the minimal form factor contains a factor which can grow doubly-exponentially as a function of rapidity differences. 

In this paper we have focused on one of the simplest and best-studied massive IQFTs, the Ising field theory. We have constructed the deformed form factors of the standard fields 
$\{\Theta (\varepsilon), \sigma, \mu\}$ and shown that they typically factorise into a function which is exactly the known, undeformed form factor and a function which depends on the deformation parameters. The latter function itself factorises into an oscillatory part and an exponential part, both of which depend on the deformation parameters. As we have shown in \cite{PRL} this structure extends beyond the Ising model, to all IQFTs with diagonal scattering. The only exception to this factorisation property seems to occur for the field $\Theta$ in the Ising model. In this case higher particle form factors are vanishing in the underformed theory but not in the deformed one. Nonetheless, even in this case we find factorisation into a function of the deformation parameters and a function that does not depend on those parameters. As expected, our solutions reduce to the known form factors  when all deformation parameters vanish. 

Starting from form factor solutions for local and semi-local fields depending on a set of continuous deformation parameters we can now expand the correlation functions and examine their properties. There is an infinite set of distinct deformations we can study, as many as the parameter choices we can make, but in essence the leading behaviours are determined by the term associated with the highest spin $s^*$ which is involved in the scattering matrix (\ref{2}) and minimal form factor (\ref{varphi2}), both characterised by the coupling $\alpha_{s^*}$. 
\bi
\item If the parameter $\alpha_{s^*}>0$ correlators are divergent at short distances, i.e. in the UV limit. In this case the UV limit is described by fundamental excitations of finite, positive length, and the divergence reflects the fact that the deep UV limit can no longer be probed. It is interesting to mention that also in the TBA context, for the case of a Gibbs ensemble, taking $\alpha_{s^*}>0$ leads to convergence problems, in this case affecting the recursive procedure typically used to solve the TBA equations. 

\item If $\alpha_{s^*}>0$ correlators are generally also divergent at large distances, unless a rapidity cut-off is introduced. The physical idea behind this is that if distances are large but momentum is also large, the finite-length of the excitations again comes into focus and correlators are divergent. However, for sufficiently low rapidities, convergence can be preserved. There is a natural rapidity cut-off which, interestingly, is described by the Lambert $W$-function, a function that is typically encountered in the context of delay equations \cite{waterloo} and also in the GHD description of $\TTb$ perturbed theories \cite{Cardy:2020olv} and ${\mathrm{JT}}$-gravity \cite{Dubovsky:2017cnj,MannOh}.

\item If the parameter $\alpha_{s^*}<0$ then the form factor expansion of correlation functions is rapidly convergent -- faster than in the unperturbed model -- both for large and short distances. The form factor expansion suggests that for short distances two-point functions still scale as power laws $r^{-4 z^{\mathcal{O}}(\bal)}$, just like in critical models. The exponents $z^{\mathcal{O}}(\bal)$ are smaller compared to the unperturbed ones $z^{\mathcal{O}}(\bal)\leq z^{\mathcal{O}}(\bol)$ and decay as $-\alpha_{s^*}$ grows
\beq 
\lim_{\alpha_{s*} \rightarrow -\infty} z_n^{\mathcal{O}}(\bal)=0\,.
\eeq 
Here $z^{\mathcal{O}}(\bal):=\sum_n z_n^{\mathcal{O}}(\bal)$ and $z_n^{\mathcal{O}}(\bal)$ is the $n$-particle contribution to the exponent in the form factor series. 
For the field $\mu$, this behaviour is found both from the short-distance scaling of correlation functions and from the $\Delta$-sum rule \cite{DSC} in the two-particle approximation. It is also possible to employ Zamolodchikov's $c$-theorem to see that the ``perturbed" $c$-function is a rapidly decreasing function of $\bal$.
\item The physical picture suggested by these results is that for $\alpha_{s^*}<0$ the perturbed IQFT flows between a UV theory, characterised by finite $\bal$-dependent critical exponents and and ``effective" central charge, and a ``trivial" IR fixed point dictated by the presence of a mass scale. 
\ei 
There are of course many open questions and possible extensions of this work.

First, although we have proposed an intuitive interpretation in the paper, the precise role and meaning of the minimal form factor ``CDD" factor \eqref{betas} remains to be better understood. 

Second, and connected to the previous point, given how reliant all our results are on the minimal form factor, it is highly desirable to find an independent method to derive  \eqref{varphi2} and \eqref{betas} and to show that these are the only possible minimal solutions. A possible approach is perturbation theory. Since the modification of the minimal form factor is universal, it would be sufficient to derive it for one simple theory, such as the Ising field theory.  

Third, in the convergent regime, it is important to understand what the form factor expansion of correlation functions is telling us about the underlying theory. For the few examples analysed here we find power-law scaling at short-distances and, although this is not evident from the figures, there is also an underlying oscillatory behaviour of the  exponents as functions of $\alpha$ (i.e. they are not monotonic functions of $\alpha$).
It is known from the existing literature that the UV theory cannot be a CFT in the standard sense, thus, how should we interpret the scaling behaviours that we have found? A related  open question is what the $\Delta$'s and $c$ that we obtain from the $\Delta$-sum rule and Zamolodchikov's $c$-theorem actually represent for a $\TTb$-deformed theory. Even though the two-particle estimate of $F_\mu(\alpha)$ gives a decreasing function that flows from $1/16$ to zero and which shares some properties with the exponent $z_2^\mu(\alpha)$ it is most likely that these are two different functions. 

For instance, if we were to add further terms in the form factor expansion of the correlator $\bra \mu(0)\Theta(r)\ket$, there is no guarantee those terms will be positive or even real for $\bal \neq \bol$, since the form factors involved contain square roots of functions that are not positive-definite. Thus the function resulting from the $\Delta$-sum rule is almost surely not the same as the power in the power-law short-distance scaling of the two-point function $\bra \mu(0)\mu(r)\ket$ where all terms must be positive and real. An interesting feature of the Ising model is that the form factors of all fields in the unperturbed theory are such that cumulants diverge for large rapidities. This leads to an interesting feature, namely that the functions $z_2^\mu(\alpha), z_2^{\varepsilon}(\alpha)$ have a sharp discontinuity exactly at $\alpha=0$. In other words, for the Ising field theory, not only are the $\alpha>0$ and $\alpha<0$ regimes extremely different but, from the form factor viewpoint, the point $\alpha=0$ is also distinct from the limit $\alpha \rightarrow 0$ of either of them.

Given the finite length of elementary excitations, one would expect that the $\Delta$-sum rule and $c$-theorem, formulae where there is integration over 'all' length scales should be modified. Indeed, our results for the $c$-function, while no doubt encapsulate some interesting physics, also show that this function is extremely different from its counterpart in the TBA context. In fact, we know that for $\alpha<0$ the TBA $c$-function becomes complex in the UV limit, a fact that is directly related to the existence of a Hagedorn transition \cite{Hagedorn:1965st}. The standard $c$-theorem, does not allow for such a behaviour, simply because the two-point function of the trace of the stress-energy tensor is positive-definite by construction. 

Fourth, another interesting problem we have already started analysing is the extension of this derivation to the branch point twist fields that play such a prominent role in the context of entanglement \cite{entropy, SymResFF}. 

Fifth, it would be nice to connect this program to existing work in CFT. It should be possible by carrying a massless limit of correlation functions/form factors to recover previous results, also in the context of entanglement.  

Finally,
 we would like to  study other (interacting) IQFTs. The beginning of such a study has been provided in \cite{PRL}. 

\medskip 

\noindent {\bf Acknowledgments:} The authors thank John Donahue, Benjamin Doyon, Fedor Smirnov, Roberto Tateo and Alexander Zamolodchikov for useful discussions. Olalla A. Castro-Alvaredo thanks EPSRC for financial support under Small Grant EP/W007045/1. Fabio Sailis is grateful for his PhD Studentship which is funded by City, University of London. The work of Stefano Negro is partially supported by the NSF grant PHY-2210349 and by the Simons Collaboration on Confinement and QCD Strings. This project was partly inspired by a meeting at the Kavli Institute for Theoretical Physics (Santa Barbara) in September 2022. Olalla A. Castro-Alvaredo and Stefano Negro thank the Institute for financial support from the National Science Foundation under Grant No. NSF PHY-1748958, and hospitality during the conference ``Talking Integrability: Spins, Fields and  Strings", August 29-September 1 (2022) and the related extended program on ``Integrability in String, Field and Condensed Matter Theory", August 22-October 14 (2022).

\appendix
\section{Correlation Functions and Cumulants for more General $\TTb$ Perturbations}

In this Appendix we discuss briefly two cases:  the case when $\alpha\neq 0$ and $\alpha_s\neq 0$ for some $s\neq 1$ (that is, there are two free parameters, a generalised $\mathrm{T}\overline{\mathrm{T}}$-perturbation involving $\mathrm{T}\overline{\mathrm{T}}$ and a higher spin $s$ operator), and  the case when only a spin $s$ term is present $\alpha_s \neq 0$ but also the parameter $\beta_s\neq 0$ with all other $\alpha$s and $\beta$s equal zero. 

\subsection{Cut-off for a Generalised $\mathrm{T}\overline{\mathrm{T}}$ Perturbation: $\alpha,\alpha_s\neq 0$}
Consider the case when two parameters in the vector $\bal$ are non-zero. We choose $\alpha_1=\alpha$ and $\alpha_s$ with $s>1$. The form of the cumulants is just a special case of the general formulae \eqref{cumu} and \eqref{cuthe}. The analysis is rather similar to that of subsection 4.2.1.

For $\alpha_s>0$ and any values of $\alpha$ (positive or negative), we can again find (for large $r$) a rapidity cut-off below which convergence can be maintained. In this case, the condition \eqref{comp} is replaced by 
\beq 
2 r \cosh \frac{x}{2} > \frac{\alpha x}{\pi} \sinh x + \frac{\alpha_s x}{\pi} \sinh (s x),
\eeq 
and for large $x$ it is the term $\sinh(s x)$ that dominates, giving the following equation for the cut-off 
\beq 
\frac{2\pi r }{\alpha_s}= {x} e^{(s-\frac{1}{2})x}\,.
\label{711}
\eeq 
which has solution 
\beq 
\Lambda=\frac{1}{s-\frac{1}{2}} W_0\left(\frac{2\pi r (s-\frac{1}{2})}{\alpha_s}\right)\,.
\label{lessbasic}
\eeq 
In other words, the contribution of the parameter $\alpha_s$ with larger spin dominates over the lower one $\alpha$ and the above result holds whether $\alpha$ is positive or negative.

\subsection{Cut-off for a Generalised $\mathrm{T}\overline{\mathrm{T}}$ Perturbation: $\alpha_s,\beta_s \neq 0$}
We now take a look at the cut-off for the two-particle cumulant in the case when only one higher spin charge is present $\alpha_s\neq 0$ but we allow for the parameter $\beta_s$ in \eqref{betas} to be non-zero as well. 

In this case, convergence of the integral in the two-particle cumulant is achieved for values of rapidity $x$ satisfying 
\beq 
2 r \cosh \frac{x}{2} > \frac{\alpha_s x}{\pi} \sinh (s x) + \beta_s \cosh (s x)\,.
\eeq 
For large rapidities, the inequality above can be approximated by
\beq 
\frac{2r \pi}{\alpha_s} > (x  + \frac{\pi \beta_s}{\alpha_s})e^{x(s-\frac{1}{2})}\,.
\eeq 
The cut-off is the solution of this equation when inequality is replaced by equality and terms are rearranged so generate an equation whose solution may be expressed in terms of Lambert's function. This can be done by multiplying both sides of the equation by a factor
\beq 
\frac{2r \pi}{\alpha_s}\left(s-\frac{1}{2}\right) e^{\frac{\pi \beta_s}{\alpha_s}\left(s-\frac{1}{2}\right)}  = \left(s-\frac{1}{2}\right)\left(x  + \frac{\pi \beta_s}{\alpha_s}\right)e^{(x+\frac{\pi \beta_s}{\alpha_s})\left(s-\frac{1}{2}\right)} \,,
\eeq
so that the solution is
\beq 
\left(s-\frac{1}{2}\right)\left(\Lambda  + \frac{\pi \beta_s}{\alpha_s}\right)=W_0\left( \frac{2r \pi}{\alpha_s}\left(s-\frac{1}{2}\right) e^{\frac{\pi \beta_s}{\alpha_s}\left(s-\frac{1}{2}\right)}  \right)\,,
\eeq 
or 
\beq 
\Lambda  =\frac{1}{s-\frac{1}{2}}W_0\left( \frac{2r \pi}{\alpha_s}\left(s-\frac{1}{2}\right) e^{\frac{\pi \beta_s}{\alpha_s}\left(s-\frac{1}{2}\right)}  \right)- \frac{\pi \beta_s}{\alpha_s}\,,
\eeq 
which reduces to \eqref{cutoff} for $s=1, \beta_s=0, \alpha_s=\alpha$ and to \eqref{lessbasic} for $\beta_s=0$.

%\bibliography{bibliography}

\end{document}